\documentclass[preprintnumbers,amsmath,amssymb,floatfix,
11pt,prd,onecolumn,superscriptaddress,nofootinbib,
longbibliography]{revtex4-2}

\usepackage[utf8]{inputenc}
\usepackage{bm}
\usepackage{amsfonts}
\usepackage{latexsym}
\usepackage{graphicx}
\usepackage{palatino}
\usepackage{mathpazo}
\usepackage{textcomp}
\usepackage{physics}
\usepackage{mathrsfs}
\usepackage{booktabs}
\usepackage{dcolumn}
\usepackage{ragged2e}
\usepackage{hyperref}
\usepackage{tensor}
\usepackage{multirow}
\usepackage{makecell}
\usepackage[version=3]{mhchem}
\usepackage{xcolor}
\usepackage{enumitem}
\usepackage{orcidlink}
\usepackage{epsfig}
\usepackage{subfigure}
\usepackage{commath}
\usepackage{caption}

\makeatletter
\renewcommand\frontmatter@abstractheading{%
  \vspace{0.2cm}
  {\centering \small\textbf{\MakeUppercase{\abstractname}}\par}
  \vspace{0.2cm} 
}
\makeatother

\usepackage{hyperref}
\hypersetup{
    colorlinks=true,
    citecolor=blue,
    linkcolor=blue,
    filecolor=blue,
    urlcolor=blue
}

\begin{document}

\title{\Large Late-time dark energy dynamics in $f(Q)$ gravity: A data-driven analysis}

\author{Kshetrimayum Govind Singh\orcidlink{0009-0009-0049-0197}}
\email{govindksh@gmail.com}
\affiliation{Department of Mathematics, Manipur University, Canchipur, Imphal, 795003, Manipur, India}%

\author{Kangujam Priyokumar Singh\orcidlink{0000-0002-8784-4091}}
\email{pk\_mathematics@yahoo.co.in}
\affiliation{Department of Mathematics, Manipur University, Canchipur, Imphal, 795003, Manipur, India}

\begin{abstract}
The increasing precision and complementarity of late-time cosmological observations provide new opportunities to test modified gravity against the observed expansion history of the universe. The combination of DESI DR2 BAO measurements with cosmic chronometers (CC) and independent Type Ia supernova compilations provides a data-driven test of cosmic acceleration. Motivated by this progress, we investigate late-time dark energy dynamics in symmetric teleparallel $f(Q)$ gravity by adopting the power-law form $f(Q)=\beta Q^{m+1}$ and characterizing cosmic evolution through the effective equation of state (EoS) parameter. The model is constrained using CC, DESI DR2 BAO and three independent Type Ia supernova compilations, namely Pantheon+, DES-Dovekie and Union3. The inferred Hubble constant values are mutually consistent, yielding $H_0=68.0^{+2.7}_{-3.3}$, $68.2^{+2.7}_{-3.3}$ and $67.4^{+2.7}_{-3.3}\,\mathrm{km\,s^{-1}\,Mpc^{-1}}$, respectively. Statistically, the proposed $f(Q)$ model yields lower $\chi^2_{\min}$ values and is favored by the AIC for all dataset combinations, whereas the BIC favors $\Lambda$CDM for the CC + DESI DR2 + Pantheon+ and CC + DESI DR2 + DES-Dovekie combinations and indicates no meaningful preference between the two models for the CC + DESI DR2 + Union3 dataset combinations. The reconstructed effective EoS characterizes the evolution of the cosmic fluid, while the deceleration parameter indicates the current accelerated expansion of the universe. Higher-order cosmographic parameters and the Statefinder and $Om(z)$ diagnostics provide complementary insights into the evolution of the cosmic expansion, while the Statefinder trajectories approach the $\Lambda$CDM fixed point in the far future. Overall, the power-law $f(Q)$ model provides a competitive description of the late-time expansion history and the observed cosmic acceleration.

\vspace{0.3cm} 
\end{abstract}

\maketitle


\section{Introduction}\label{sec:3.01}
The field of observational cosmology has witnessed remarkable progress in recent years, driven by increasingly precise measurements of the cosmic expansion history over an extended redshift range. The discovery of late-time cosmic acceleration through observations of distant Type Ia supernovae represented a fundamental development in modern cosmology~\citep{Riess_1998,Perlmutter_1999,Riess_2004}. 
This accelerated expansion is commonly attributed to dark energy, an unknown component with negative effective pressure that dominates the late-time energy content of the universe. Dark energy is commonly characterized through its equation of state (EoS) parameter $(\omega)$, with the simplest realization being the cosmological constant $\Lambda$, for which $\omega=-1$. Since the discovery of cosmic acceleration, independent cosmological probes have substantially improved our understanding of the background evolution of the universe. In particular, observations of the cosmic microwave background (CMB)~\citep{Caldwell_2004}, baryon acoustic oscillations (BAO)~\citep{Eisenstein_2005,Percival_2010}, large-scale structure (LSS)~\citep{Koivisto_2006,Daniel_2008}, together with other observational data~\citep{Spergel_2003,Abdul_2025}, have provided complementary information on the expansion history and the underlying dynamics of the universe. The combined observational picture has established the $\Lambda$ cold dark matter ($\Lambda$CDM) framework as the reference model for describing the large-scale evolution of the universe. Within general relativity (GR)~\citep{Einstein_1916}, the accelerated expansion is attributed to the cosmological constant, which provides an excellent description of a broad range of cosmological observations. Nevertheless, interpreting dark energy as a cosmological constant raises major conceptual challenges, most notably the fine-tuning~\citep{Weinberg_1989} and cosmic coincidence problems~\citep{Dalal_2001}. Moreover, the tension between independent measurements of the present-day Hubble expansion rate, $H_0$, has intensified interest in the nature and dynamics of dark energy and has motivated the exploration of alternatives to the standard cosmological scenario~\citep{Verde_2019,Valentino_2021}.

These challenges have stimulated considerable interest in exploring the nature of dark energy and the physical mechanism responsible for the late-time cosmic acceleration. Broadly, the origin of the accelerated expansion can be investigated from two complementary perspectives. The first is to retain GR as the underlying theory of gravity and introduce an additional component in the matter-energy sector capable of driving the accelerated expansion. This approach encompasses a wide range of dark energy models, including the cosmological constant, quintessence, $k$-essence, tachyon, phantom and Chaplygin-gas models~\citep{Copeland_2006,Armendariz_2001,Bagla_2003,Bento_2002,Debnath_2004}. The second perspective is to modify the gravitational sector itself, thereby seeking to explain the observed acceleration through an extension of the geometric description of gravity rather than through the introduction of a new dark energy component. This latter approach gives rise to a broad class of modified-gravity theories in which the gravitational dynamics are generalized beyond the framework of GR. Such theories provide an alternative avenue for investigating the late-time expansion history and the underlying dynamics of cosmic acceleration. Among these theories, $f(R)$ gravity~\citep{Buchdahl_1970,Starobinsky_1980} extends the Einstein-Hilbert description by replacing the Ricci scalar $R$ in the gravitational action with a general function of $R$. More general formulations have also been constructed by introducing direct interactions between geometry and matter, as in $f(R,T)$ and $f(R,L_m)$ gravity~\citep{Harko_2011,Harko_2010}, where $T$ represents the trace of the energy--momentum tensor and $L_m$ denotes the matter Lagrangian density. Modified descriptions can alternatively be formulated using torsion or non-metricity as the fundamental geometric quantities. For example, $f(T)$ gravity~\citep{Ferraro_2007} is based on the torsion scalar $T$, whereas $f(Q)$ gravity~\citep{Jimenez_2018} employs the non-metricity scalar $Q$ within the framework of symmetric teleparallel gravity. In the latter formulation, gravity is described geometrically through non-metricity rather than spacetime curvature. More specifically, the symmetric teleparallel construction is characterized by vanishing curvature and torsion, with the non-metricity sector carrying the relevant gravitational degrees of freedom. This framework retains second-order field equations and can generate effective accelerated expansion for suitable choices of $f(Q)$, making it an attractive approach for investigating cosmic evolution. 

The development of $f(Q)$ gravity has opened another geometric route for describing gravitational dynamics. The formulation introduced by~\citet{Jimenez_2018} constructs the symmetric teleparallel equivalent of gravity in terms of the non-metricity scalar $Q$, thereby providing the geometric foundation for the $f(Q)$ framework. A detailed account of the theory and its broader applications has subsequently been presented by~\citet{Heisenberg_2024}. Cosmological investigations have extended beyond spatially flat backgrounds as well. For example, a dynamical-systems analysis by~\citet{Shabani_2024} demonstrated that modified $f(Q)$ models admit a variety of cosmological trajectories, encompassing matter dominated and dark energy dominated regimes as well as accelerating solutions associated with spatial curvature. Constraints arising from the energy conditions have also been studied in this framework.~\citet{Mandal_2020} examined the relevant energy conditions and obtained restrictions on the model parameters that are consistent with an accelerating cosmic expansion. Observational aspects of the theory have likewise received increasing attention.~\citet{Solanki_2021} considered a modified $f(Q)$ scenario containing bulk-viscous matter and confronted it with $H(z)$, Type Ia supernova and BAO data. Their analysis demonstrated that the model permits a cosmological history in which the universe evolves from an earlier decelerating state to the currently observed accelerated phase. More recently, we investigated the observational viability of $f(Q)$ gravity through a logarithmic $Om(z)$ framework
~\citep{Singh_2026Observational}. A number of subsequent investigations have further examined the cosmological implications and phenomenological aspects of $f(Q)$ gravity~\citep{Arora_2026,Nashed_2026,Paliathanasis_2026,Kolhatkar_2026,Mazumdar_2026,Chakraborty_2025,Dubey_2025,Yadav_2024,Narawade_2023,Koussour_2022,Lymperis_2022,Lazkoz_2019}.

A phenomenological description of late-time cosmic expansion can be constructed through an effective EoS parametrization. Such a parametrization characterizes the redshift evolution of the effective cosmic fluid without specifying its underlying physical origin. One such parametrization was proposed by~\citet{Mukherjee_2016}, given by
$$\omega(z)=
-\frac{1}{1+\gamma(1+z)^n},$$
where $\gamma$ and $n$ are model parameters. For $\gamma>0$ and $n>0$, the EoS parameter approaches zero at high redshift and tends to \(-1\) in the asymptotic future. This behavior describes an evolution of the effective EoS from a matter-like value at early times toward a cosmological-constant-like value in the far future. The parametrization thus provides a useful phenomenological framework for investigating the evolution of the effective cosmic fluid in modified gravity.

In the present work, we employ this EoS parametrization within the framework of $f(Q)$ gravity to investigate the late-time expansion history of the universe using recent and high-precision cosmological observations.
In particular, we combine the BAO measurements from the Dark Energy Spectroscopic Instrument (DESI) Data Release 2 (DR2)~\citep{Abdul_2025,Lodha_2025} with cosmic chronometer (CC) data and three complementary Type Ia supernova (SNe Ia) compilations, namely Pantheon+, DES-Dovekie and Union3~\citep{Brout_2022,Popovic_2026,Rubin_2025}. The model parameters are constrained through a Markov chain Monte Carlo (MCMC) analysis~\citep{Foreman_2013}, allowing us to examine how different combinations of observational datasets affect the inferred cosmological parameters and the reconstructed expansion history. We assess the consistency of the resulting $f(Q)$ cosmology with the standard $\Lambda$CDM model and investigate the evolution of the effective EoS, the deceleration parameter and the transition from decelerated to accelerated expansion. Furthermore, we employ the $Om(z)$ and Statefinder diagnostics to provide complementary tests of the dynamical behavior of the model and to identify possible departures from the $\Lambda$CDM scenario. By combining a phenomenological effective EoS with the geometric framework of $f(Q)$ gravity and current observational constraints, this study provides a data-driven assessment of the model's ability to reproduce the observed late-time expansion history of the universe.

The paper is organized as follows. Section~\ref{sec:3.02} provides a brief review of the theoretical framework of $f(Q)$ gravity. Section~\ref{sec:3.03} presents the cosmological framework with the parametrized effective equation of state and derives the corresponding Hubble parameter. The observational datasets and statistical methodology employed for constraining the model parameters are described in Section~\ref{sec:3.04}. Section~\ref{sec:3.05} is devoted to the cosmographic analysis, including the present-day cosmographic parameters and the deceleration-to-acceleration transition. 
Section~\ref{sec:3.06} presents the observationally constrained evolution of the effective energy density, pressure and equation of state parameter and discusses their implications for late-time cosmic acceleration. The observationally constrained cosmic dynamics are further examined through the statefinder and $Om(z)$ diagnostics in Section~\ref{sec:3.07}. Finally, the main results of the analysis and their cosmological implications are summarized in Section~\ref{sec:3.08}.

\section{Brief review of $f(Q)$ gravity}\label{sec:3.02}
The action for $f(Q)$ gravity is given by~\citep{Jimenez_2018}:
\begin{equation}\label{eq:3.01}
S = \int\left[\frac{1}{2} f(Q) + L_{m} \right] \sqrt{-g} d^{4}x , 
\end{equation}
where $L_m$ represents the matter Lagrangian density, $f(Q)$ is an arbitrary function of the non-metricity scalar $Q$ and $g$ denotes the determinant of the metric tensor $g_{\mu\nu}$.

The non-metricity tensor and its two traces are defined as
\begin{equation}\label{eq:3.02}
Q_{\gamma\mu\nu} = \nabla_\gamma g_{\mu\nu},
\end{equation}
\begin{equation}\label{eq:3.03}
Q_\gamma = Q_{\gamma\ \mu}^{\ \mu}, \qquad \tilde{Q}_\gamma = Q^\mu_{\ \gamma\mu}.
\end{equation}
Moreover, the non-metricity conjugate (superpotential tensor) is given by
\begin{equation}\label{eq:3.04}
4 P^\gamma{}_{\mu\nu} = -Q^\gamma{}_{\mu\nu} + 2 Q_{(\mu}{}^\gamma{}_{\nu)} - Q^\gamma g_{\mu\nu} - \tilde{Q}^\gamma g_{\mu\nu} - \delta^\gamma_{(\mu} Q_{\nu)}.
\end{equation}
The non-metricity scalar is then defined as
\begin{equation}\label{eq:3.05}
Q = -Q_{\gamma\mu\nu} P^{\gamma\mu\nu}.
\end{equation}
The energy--momentum tensor associated with the matter sector is given by
\begin{equation}\label{eq:3.06}
T_{\mu\nu} = -\frac{2}{\sqrt{-g}} \frac{\delta (\sqrt{-g}\,L_m)}{\delta g^{\mu\nu}}.
\end{equation}
Varying the action~\eqref{eq:3.01} with respect to the metric tensor leads to the modified field equations:
\begin{equation}\label{eq:3.07}
\frac{2}{\sqrt{-g}} \nabla_{\gamma} \left( \sqrt{-g} f_{Q} P^{\gamma}{}_{\mu\nu} \right) + \frac{1}{2} g_{\mu\nu} f
 + f_{Q} \left( P_{\mu\gamma\alpha} Q_{\nu}{}^{\gamma\alpha} - 2 Q_{\gamma\alpha\mu} P^{\gamma\alpha}{}_{\nu} \right) = -T_{\mu\nu},
\end{equation}
where $f_{Q} = \frac{df}{dQ}$. 
Furthermore, varying the action~\eqref{eq:3.01} with respect to the connection yields

\begin{equation}\label{eq:3.08}
\nabla_{\mu} \nabla_{\gamma} \left( \sqrt{-g} f_{Q} P^{\gamma}{}_{\mu\nu} \right) = 0.
\end{equation}

We consider a spatially flat, homogeneous and isotropic Friedmann--Lemaître--Robertson--Walker (FLRW) spacetime~\citep{Ryden_2003},
\begin{equation}\label{eq:3.09}
ds^{2} = -dt^{2} + a^{2}(t) \left( dx^{2} + dy^{2}+dz^{2}\right),
\end{equation}
where $a(t)$ represents the scale factor of the universe. For the FLRW spacetime, the non-metricity scalar becomes
\begin{equation}\label{eq:3.10}
Q = 6H^{2},
\end{equation}
where $H=\frac{\dot{a}}{a}$ denotes the Hubble parameter.
The energy--momentum tensor of a perfect fluid is given by
\begin{equation}\label{eq:3.11}
T_{\mu\nu} = (\rho + p) u_{\mu} u_{\nu} + p g_{\mu\nu},
\end{equation}
where $\rho$ and $p$ are the energy density and pressure, respectively and $u_{\mu}$ denotes the four-velocity vector of the perfect fluid which satisfies $u_{\mu} u^{\mu} = -1$. Upon substituting Eqs.~\eqref{eq:3.09} and~\eqref{eq:3.11} into Eq.~\eqref{eq:3.07}, the modified Friedmann equations governing the cosmological dynamics in $f(Q)$ gravity are obtained as

\begin{equation}\label{eq:3.12}
    6 f_Q H^2-\frac{1}{2} f=\rho,
\end{equation}
\begin{equation}\label{eq:3.13}
  (12 f_{QQ}H^2+f_Q)\dot{H}=-\frac{1}{2}(\rho+p).
\end{equation}
 It is worth noting that General Relativity is recovered for the linear choice $f(Q)=Q$. In this limit, the modified Friedmann equations reduce to the standard Friedmann equations,
\begin{equation}\label{eq:3.14}
3H^2=\rho,
\end{equation}

\begin{equation}\label{eq:3.15}
2\dot{H}+3H^2=-p.
\end{equation}

In the present work, we consider a spatially homogeneous and isotropic universe comprising pressureless matter and a dark energy sector. Since our analysis focuses on the late-time evolution of the universe, the radiation component is neglected. The matter component is assumed to be dust with vanishing pressure ($p_m=0$), whereas dark energy is characterized by a negative pressure that drives the observed late-time accelerated expansion of the universe. Accordingly, the total energy density and pressure are expressed as
$\rho=\rho_m+\rho_{\rm DE}, \;
p=p_{\rm DE}$. Rather than assigning the phenomenological EoS directly to the dark energy component, we characterize the background cosmic dynamics through the effective EoS parameter of the total cosmic fluid. It is defined as the ratio of the total pressure to the total energy density.
Using the modified Friedmann Eqs.~\eqref{eq:3.12} and~\eqref{eq:3.13}, this effective EoS parameter can be written as
\begin{equation}\label{eq:3.16}
\omega=\frac{p}{\rho}
=\frac{f-12f_QH^2-4\left(12f_{QQ}H^2+f_Q\right)\dot{H}}{12f_QH^2-f}.
\end{equation}

\section{Cosmological framework with a parametrized effective equation of state}\label{sec:3.03}
Parametrizations of the EoS parameter provide a convenient phenomenological framework for describing the dynamical evolution of the cosmic fluid responsible for the late-time accelerated expansion of the universe. In the absence of a fundamental theory that uniquely determines the redshift dependence of the EoS parameter, phenomenological parametrizations have been widely employed to investigate possible deviations from the standard $\Lambda$CDM cosmology and to confront theoretical models with observational data. Throughout this work, the cosmological evolution is described in terms of the redshift $z$. Expressing cosmological quantities as functions of redshift allows a direct comparison between theoretical predictions and observational measurements. A variety of EoS parametrizations have been proposed in the literature, including the Chevallier--Polarski--Linder (CPL)~\citep{CHEVALLIER_2001,Linder_2003}, Linear~\citep{Weller_2002}, Logarithmic~\citep{Efstathiou_1999}, Jassal--Bagla--Padmanabhan (JBP)~\citep{Jassal_2005} and Barboza--Alcaniz (BA)~\citep{Barboza_2008} forms, among others. These parametrizations have been extensively constrained using different combinations of cosmological observations and have proven useful in characterizing the dynamical evolution of the cosmic expansion history. 

Motivated by the need for a simple yet flexible description of the cosmic expansion history, we consider the following effective EoS parametrization~\citep{Mukherjee_2016}:
\begin{equation}\label{eq:3.17}
\omega(z)= -\frac{1}{1+\gamma (1+z)^n},
\end{equation}
where $\gamma$ and $n$ are model parameters. For positive values of $\gamma$ and $n$, the value of $\omega$ tends to zero at a high redshift $z$, while its present value at $z=0$ is determined by $\gamma$, given by 
$\omega_{0}= -\frac{1}{1+\gamma}$. Furthermore, the expression for $\omega$ in Eq.~\eqref{eq:3.17} shows that a positive value of the model parameter $\gamma$ always sets a lower bound on the value of the effective EoS. This confines the effective EoS to the non-phantom regime ($-1<\omega(z)<0$), thereby excluding phantom behavior by construction rather than relying on independent observational constraints.

In order to investigate the cosmological implications of $f(Q)$ gravity, we consider the power-law model
\begin{equation}\label{eq:3.18}
f(Q)=\beta Q^{m+1},
\end{equation}
where $\beta$ and $m$ are model parameters~\citep{Harko_2018}. The corresponding derivatives are $f_Q=\beta (m+1)Q^m$
and $f_{QQ}=\beta (m+1)mQ^{m-1}$. Substituting these expressions into Eqs.~\eqref{eq:3.12} and~\eqref{eq:3.13}, we obtain the energy density and pressure, respectively, as
\begin{equation}\label{eq:3.19}
\rho=\frac{\beta}{2}(2m+1)(6H^2)^{m+1},
\end{equation}
and
\begin{equation}\label{eq:3.20}
p=-\beta(2m+1)(6H^2)^m\left[2(m+1)\dot{H}+3H^2\right].
\end{equation}

To assess the basic theoretical viability of the adopted model, we examine the positivity of $f_Q$, $f_Q+2Qf_{QQ}$ and the effective energy density. For the power-law model, the second quantity takes the form $f_Q+2Qf_{QQ}=\beta(m+1)(2m+1)Q^m.$ Adopting the convention $Q=6H^2>0$ and assuming $\beta>0$, the condition $f_Q>0$ requires $m>-1$, whereas $f_Q+2Qf_{QQ}>0$ is satisfied for $m<-1$ or $m>-1/2$. Furthermore, the positivity of the energy density in Eq.~\eqref{eq:3.19} requires $m>-1/2$. The intersection of these constraints jointly restricts the parameter space to $m>-1/2$. While these positivity conditions provide a basic consistency check for the model, they are insufficient on their own to establish complete perturbative stability.

Using $1+z=\frac{a_0}{a}$ with $a_0=1$~(the present-day value of the scale factor), the time and redshift derivatives are related by
$\frac{d}{dt}=-(1+z)H(z)\frac{d}{dz}$,
which immediately yields

\begin{equation}\label{eq:3.21}
\dot{H}=-(1+z)H(z)H'(z),
\end{equation}
where a prime denotes a derivative with respect to the redshift $z$.

Using the above relation in Eq.~\eqref{eq:3.16}, together with the expressions for the energy density and pressure, the effective EoS can be written as,
\begin{equation}\label{eq:3.22}
\omega(z)
=-1+\frac{2(m+1)}{3}(1+z)\frac{H'(z)}{H(z)}.
\end{equation}
Using Eqs.~\eqref{eq:3.22} and~\eqref{eq:3.17}, we obtain
\begin{equation}\label{eq:3.23}
\frac{H'(z)}{H(z)}
=\frac{3\gamma(1+z)^{n-1}}{2(m+1)\left[1+\gamma(1+z)^n\right]}.
\end{equation}

Integrating Eq.~\eqref{eq:3.23}, the Hubble parameter as a function of redshift is obtained as
\begin{equation}\label{eq:3.24}
H(z)=H_0\left[\frac{1+\gamma(1+z)^n}{1+\gamma}
\right]^{\frac{3}{2n(m+1)}}.
\end{equation}
where $H_0\equiv H(z=0)$ denotes the present-day value of the Hubble parameter.

The reconstructed Hubble parameter in Eq.~\eqref{eq:3.24} depends on $H_0$, $\gamma$, $n$ and $m$, but is independent of the coefficient $\beta$. To determine $\beta$, we impose the present-day normalization condition $\rho_0=3H_0^2$, in units where $8\pi G=1$. Evaluating Eq.~\eqref{eq:3.19} at $z=0$, where $H(z=0)=H_0$, gives
\begin{equation}\label{eq:3.25}
3H_0^2=\frac{\beta}{2}(2m+1)(6H_0^2)^{m+1}.
\end{equation}
Solving for $\beta$, we obtain
\begin{equation}\label{eq:3.26}
\beta=\frac{1}{(2m+1)(6H_0^2)^m}.
\end{equation}
Thus, under the adopted normalization, $\beta$ is a derived quantity determined by $H_0$ and $m$, rather than an independently estimated parameter. For $m=0$, this normalization yields $\beta=1$, giving $f(Q)=Q$, which recovers the standard General Relativity. Since $\beta$ does not enter the reconstructed Hubble parameter in Eq.~\eqref{eq:3.24}, it is not independently constrained by the background expansion data within this reconstruction. The corresponding energy density and pressure are evaluated using the normalized value of $\beta$.

For \(\gamma>0\), \(n>0\) and \(m>-1\), the high-redshift limit of Eq.~\eqref{eq:3.24} gives
$H(z)\propto (1+z)^{\frac{3}{2(m+1)}}$.
Meanwhile, the EoS parametrization in Eq.~\eqref{eq:3.17} approaches \(\omega(z)\to0\) as \(z\to\infty\). Although this limiting value is conventionally associated with pressureless matter, the standard matter-dominated expansion-rate scaling, \(H(z)\propto(1+z)^{3/2}\), is recovered only for \(m=0\). Thus, the asymptotic behavior \(\omega(z)\to0\) does not, by itself, establish the recovery of the standard matter-dominated expansion history for general \(m\). The extent to which the expansion rate approaches the standard matter-dominated scaling depends on the observational constraints on \(m\).

\section{Observational data and statistical methodology}\label{sec:3.04}
In this section, we outline the observational datasets and statistical framework used to constrain the free parameters of our cosmological model. Parameter estimation is conducted within a Bayesian framework via Markov Chain Monte Carlo (MCMC) sampling. Specifically, we implement the affine-invariant ensemble sampler using the \texttt{emcee} Python package~\citep{Foreman_2013}, chosen for its efficiency in multi-dimensional parameter spaces and its robustness against non-Gaussian posteriors and parameter degeneracies.

The likelihood function is defined as
\begin{equation}\label{eq:3.27}
    \mathcal{L}(\boldsymbol{\theta}) \propto \exp\left(-\frac{\chi^2(\boldsymbol{\theta})}{2}\right),
\end{equation}
where $\chi^2$ denotes the total chi-square statistic. The parameter vector under consideration is defined as $\boldsymbol{\theta} = \{H_0, m, n, \gamma, r_{\mathrm{d}}\}$. Explicit expressions for the $\chi^2$ functions corresponding to each observational dataset are detailed in the following subsections.

We sample the parameter space by assuming flat, physically motivated uniform priors:
\begin{equation*}
    60 < H_0 < 80, \quad -0.5 < m < 3, \quad 0.01 < n < 4, \quad 0.01 < \gamma < 5, \quad 130 < r_{\mathrm{d}} < 160.
\end{equation*}

The MCMC analysis yields the full posterior probability distributions for the parameter set. In the resulting corner plot, the diagonal panels show the one-dimensional marginalized posterior distributions, yielding the median values and $1\sigma$ ($68.3\%$) credible intervals. The off-diagonal panels show the two-dimensional joint posteriors, with $1\sigma$ and $2\sigma$ confidence contours highlighting parameter correlations and degeneracies.

\subsection{Cosmic chronometers}\label{subsec:3.01}
The cosmic chronometer~(CC) method provides a direct probe of the expansion history of the universe through the differential aging of passively evolving galaxies. The method was originally proposed by Jimenez and Loeb~\citep{Jimenez_2002}, who showed that the Hubble parameter can be directly inferred from the differential evolution of cosmic time with redshift,
\begin{equation}\label{eq:3.28}
H(z)=-\frac{1}{1+z}\frac{dz}{dt}.
\end{equation}
Thus, measurements of the differential ages of passively evolving galaxies over small redshift intervals provide estimates of $H(z)$ without requiring a distance calibration. In practice, the quantity $\frac{dz}{dt}$ can be estimated from the redshift and age differences between nearby galaxy populations. Since spectroscopic redshifts can be measured with high precision, the differential-age approach allows the expansion rate to be constrained directly from the observed evolution of galaxy ages. For the present study, we adopt a compilation of 31 CC measurements covering the redshift range $0.07\leq z\leq1.965$. The measurements are collected from several independent CC analyses in the literature and the corresponding redshifts, Hubble parameter values, $1\sigma$ uncertainties and literature sources are listed in Table~\ref{tab:3.01}. 

To obtain reliable parameter constraints, we account for both statistical and systematic uncertainties. Following the methodology outlined by~\citet{Moresco_2020}, the total covariance matrix can be expressed as:
\begin{equation}\label{eq:3.29}
    \mathrm{Cov}_{ij} = \mathrm{Cov}_{ij}^{\mathrm{stat}} +  \mathrm{Cov}_{ij}^{\mathrm{young}} + \mathrm{Cov}_{ij}^{\mathrm{model}} + \mathrm{Cov}_{ij}^{\mathrm{met}},
\end{equation}
where the terms represent contributions to the covariance from statistical errors, young stellar component contamination, model dependence and stellar metallicity, respectively. The model dependent covariance contribution, $\mathrm{Cov}_{ij}^{\mathrm{model}}$, can be further decomposed as follows:
\begin{equation}\label{eq:3.30}
    \mathrm{Cov}_{ij}^{\mathrm{model}} = \mathrm{Cov}_{ij}^{\mathrm{SFH}} + \mathrm{Cov}_{ij}^{\mathrm{IMF}} + \mathrm{Cov}_{ij}^{\mathrm{st.lib.}} + \mathrm{Cov}_{ij}^{\mathrm{SPS}},
\end{equation}
representing the systematic uncertainties associated with the assumed star formation history (SFH), initial mass function (IMF), stellar libraries and stellar population synthesis (SPS) models.

Statistical errors, metallicity and SFH contributions are conventionally treated as uncorrelated diagonal elements within the published $1\sigma$ errors. In contrast, the underlying choices of SPS model, stellar library and IMF generate correlated non-diagonal covariance terms across redshift bins. To quantify these correlations, we utilize the calibration framework provided by \citet{Moresco_2020}. The covariance contribution associated with young stellar populations is assumed to be negligible for this sample of strictly passive galaxies.

We use a Gaussian likelihood to constrain the model parameters using the CC dataset, with the corresponding $\chi^2$ statistic defined as
\begin{equation}\label{eq:3.31}
\chi^2_{\mathrm{CC}} =
\Delta\mathbf{H}^{T}\mathbf{C}_{\rm CC}^{-1}
\Delta\mathbf{H},
\end{equation}
where $\Delta\mathbf{H} = \mathbf{H}_{\mathrm{obs}} - \mathbf{H}_{\mathrm{th}}(z)$ is the residual vector between the observed and theoretical Hubble parameter values and $\mathbf{C}_{\rm CC}$ is the covariance matrix accounting for both statistical and systematic uncertainties.

\begin{table}[ht]
\centering
\renewcommand{\arraystretch}{1.15}
\setlength{\tabcolsep}{9pt}
\small
\caption{\justifying The 31 cosmic chronometer (CC) data points used in this analysis. The values of $H(z)$ and their $1\sigma$ uncertainties, $\sigma_H$, are given in $\mathrm{km\,s^{-1}\,Mpc^{-1}}$.}
\label{tab:3.01}
\begin{tabular}{cccc|cccc}
\hline
$z$ & $H(z)$ & $\sigma_H$ & Ref. &
$z$ & $H(z)$ & $\sigma_H$ & Ref. \\
\hline
0.07 & 69  & 19.6 & \citep{Zhang_2014} &
0.4783 & 80.9  & 9  & \citep{Moresco_2016} \\

0.09 & 69  & 12 & \citep{Simon_2005} &
0.48 & 97  & 62 & \citep{Stern_2010} \\

0.12 & 68.6  & 26.2 & \citep{Zhang_2014} &
0.593 & 104 & 13 & \citep{Moresco_2012} \\

0.17 & 83  & 8  & \citep{Simon_2005} &
0.6797 & 92  & 8  & \citep{Moresco_2012} \\

0.1791 & 75  & 4  & \citep{Moresco_2012} &
0.7812 & 105 & 12 & \citep{Moresco_2012} \\

0.1993 & 75  & 5  & \citep{Moresco_2012} &
0.8754 & 125 & 17 & \citep{Moresco_2012} \\

0.2 & 72.9  & 29.6 & \citep{Zhang_2014} &
0.88 & 90  & 40 & \citep{Stern_2010} \\

0.27 & 77  & 14 & \citep{Simon_2005} &
0.9 & 117 & 23 & \citep{Simon_2005} \\

0.28 & 88.8  & 36.6 & \citep{Zhang_2014} &
1.037 & 154 & 20 & \citep{Moresco_2012} \\

0.3519 & 83  & 14 & \citep{Moresco_2012} &
1.3 & 168 & 17 & \citep{Simon_2005} \\

0.3802 & 83  & 13.5 & \citep{Moresco_2016} &
1.363 & 160 & 33.6 & \citep{Moresco_2015} \\

0.4 & 95  & 17 & \citep{Simon_2005} &
1.43 & 177 & 18 & \citep{Simon_2005} \\

0.4004 & 77  & 10.2 & \citep{Moresco_2016} &
1.53 & 140 & 14 & \citep{Simon_2005} \\

0.4247 & 87.1  & 11.2 & \citep{Moresco_2016} &
1.75 & 202 & 40 & \citep{Simon_2005} \\

0.4497 & 92.8  & 12.9 & \citep{Moresco_2016} &
1.965 & 186.5 & 50.4 & \citep{Moresco_2015} \\

0.47 & 89  & 34 & \citep{Ratsimbazafy_2017} &
& & & \\

\hline
\end{tabular}
\end{table}

\subsection{Baryon acoustic oscillations from DESI DR2}\label{subsec:3.02}
To map the expansion history of the universe across a broad redshift range, we include the recent BAO measurements from the DESI DR2. Derived from the first three years of observations, the DESI DR2 dataset comprises spectroscopic redshifts for over 14 million galaxies and quasars~\citep{Abdul_2025,Lodha_2025}. The BAO scale is measured from the redshift-space two-point correlation function across multiple tracers: the Bright Galaxy Survey (BGS), luminous red galaxies (LRGs), emission-line galaxies (ELGs) and quasars (QSOs). At higher redshifts, these constraints are complemented by BAO features extracted from the Lyman-$\alpha$ ($\mathrm{Ly}\alpha$) forest absorption in quasar spectra~\citep{Abdul_2025}.

The BAO scale acts as a standard ruler whose physical length is set by the comoving sound horizon at the baryon drag epoch, denoted by $r_d$. We use the complete set of 13 DESI DR2 BAO measurements reported in Table IV of~\citet{Abdul_2025}. These measurements comprise one isotropic constraint, $D_V/r_d$, from the BGS sample, together with six pairs of anisotropic measurements, $D_M/r_d$ and $D_H/r_d$, obtained from the LRG, ELG and quasar tracer samples. The measurements span an effective redshift range of $0.295 \leq z \leq 2.330$, thereby providing complementary constraints on the expansion history from low to high redshifts.

For the spatially flat FLRW geometry adopted in this work, the transverse comoving distance is defined as

\begin{equation}\label{eq:3.32}
D_M(z)=c\int_0^z\frac{dz'}{H(z')},
\end{equation}
while the Hubble distance is given by
\begin{equation}\label{eq:3.33}
D_H(z)=\frac{c}{H(z)}.
\end{equation}
The volume-averaged distance, relevant for isotropic BAO measurements, is defined as
\begin{equation}\label{eq:3.34}
D_V(z)=\left[zD_M^2(z)D_H(z)\right]^{1/3}.
\end{equation}
Since the observed BAO scale is conventionally expressed relative to the sound horizon at the drag epoch, the corresponding theoretical quantities used in the likelihood analysis are $D_M(z)/r_d$, $D_H(z)/r_d$ and $D_V(z)/r_d$~\citep{Abdul_2025}. These observables enable the DESI DR2 BAO data to constrain both the transverse and radial components of the cosmological expansion history.

Rather than fixing the sound horizon $r_d$ to a value determined by a specific early-universe model, we treat $r_d$ as a free parameter in the MCMC analysis. This choice avoids imposing an external early-universe calibration of the BAO standard ruler and allows the DESI DR2 data to constrain the expansion history without assuming a fixed value of $r_d$. The theoretical data vector, $\mathbf{D}_{\rm th}$, is constructed from the model predictions for $D_V/r_d$, $D_M/r_d$ and $D_H/r_d$ evaluated at the corresponding effective redshifts, while $\mathbf{D}_{\rm obs}$ denotes the observed BAO data vector. To account for the correlations among the DESI DR2 measurements, we use the full $13\times13$ covariance matrix, $\mathbf{C}_{\rm DESI}$. The BAO contribution to the total $\chi^2$ is therefore given by
\begin{equation}\label{eq:3.35}
\chi^2_{\rm DESI\,DR2}=\Delta\mathbf{D}^{T}\mathbf{C}_{\rm DESI}^{-1}
\Delta\mathbf{D},
\end{equation}
where $\Delta\mathbf{D} = \mathbf{D}_{\mathrm{obs}} - \mathbf{D}_{\mathrm{th}}(z)$ is the residual vector between the observed and theoretical BAO observables.

\subsection{Type Ia supernovae}\label{subsec:3.03}
We employ three complementary Type Ia supernovae (SNe Ia) compilations, namely Pantheon+, DES-Dovekie and Union3, to constrain the late-time expansion history of the universe. SNe Ia serve as standardizable candles, with their standardized apparent magnitudes enabling luminosity distances to be inferred through the magnitude--redshift relation. The broad redshift coverage and large sample sizes of these compilations help constrain the cosmological parameters.

\medskip
\noindent\textbf{Pantheon+:} The Pantheon+ compilation contains 1701 light curves corresponding to 1550 distinct SNe Ia over the redshift range $0.001 \leq z \leq 2.26$~\citep{Brout_2022,Scolnic_2022}. Compared with the original Pantheon sample, it features an increased sample size and redshift coverage, improved photometric calibration and a more comprehensive treatment of systematic uncertainties. For the present analysis, we exclude the 77 SH0ES Cepheid-calibrator supernovae and use the remaining 1624 non-calibrator SNe Ia, thereby avoiding the use of external Cepheid distance calibration. 

\medskip
\noindent\textbf{DES-Dovekie:}
The DES-Dovekie Type Ia supernova sample is a recalibrated and reanalysed version of the Dark Energy Survey five-year supernova dataset (DES-SN5YR)~\citep{Abbott_2024}, presented by~\citet{Popovic_2026}. It contains 1820 supernova light curves, including 1623 photometrically classified SNe Ia observed with the Dark Energy Camera (DECam) over the redshift range $0.10 \leq z \leq 1.14$ and 197 spectroscopically confirmed low-redshift SNe Ia from external surveys over $0.025 \leq z \leq 0.10$. The DES-Dovekie reanalysis provides improved photometric calibration and an updated treatment of systematic uncertainties.

\medskip
\noindent\textbf{Union3:} We also employ the Union3 compilation, which contains 2087 cosmologically useful SNe Ia assembled from 24 different observational datasets~\citep{Rubin_2025}. The compilation places the SNe Ia on a common distance scale using the SALT3 light-curve model for consistent standardization of their luminosities. The Union3 analysis employs the UNITY1.5 Bayesian framework, which accounts for selection effects, intrinsic dispersion, outliers and systematic uncertainties in the statistical inference. For cosmological analyses, the compilation provides 22 binned estimates of the SNe distance modulus over the redshift range $0.05 \leq z \leq 2.26$. These binned measurements provide a compact representation of the observed distance--redshift relation and are used to test the theoretical expansion history against the Union3 data.

The $\chi^2$ function for the SNe Ia data is defined as~\citep{Perivolaropoulos_2023}
\begin{equation}\label{eq:3.36}
\chi^2_{\mathrm{SNe}}=
\Delta\boldsymbol{\mu}^{\,T}
C_{\mathrm{stat+syst}}^{-1}
\Delta\boldsymbol{\mu},
\end{equation}
where $C_{\mathrm{stat+syst}}$ denotes the combined statistical and systematic covariance matrix and $\Delta\boldsymbol{\mu}$ is the vector of residuals between the observed and theoretical distance moduli, with components
\begin{equation}\label{eq:3.37}
\Delta\mu_i=\mu_{\mathrm{obs},i}-\mu_{\mathrm{th}}(z_i).
\end{equation}
The observed and theoretical distance moduli are given, respectively, by
\begin{equation}\label{eq:3.38}
\mu_{\mathrm{obs},i}=m_{B i}-M
\end{equation}
and
\begin{equation}\label{eq:3.39}
\mu_{\mathrm{th}}(z_i)=
5\log_{10}\left[\frac{d_L(z_i)}{\mathrm{Mpc}}\right]+25,
\end{equation}
where $m_{B i}$ denotes the apparent magnitude of the $i$-th SNe Ia, $M$ is the absolute magnitude and $d_L(z)$ is the luminosity distance. For a spatially flat FLRW universe, the luminosity distance is given by
\begin{equation}\label{eq:3.40}
d_L(z)=(1+z)c\int_0^z\frac{dz'}{H(z')},
\end{equation}
where $c$ is the speed of light and $H(z)$ is the Hubble parameter. For all three SNe Ia datasets, the absolute magnitude $M$ is treated as a nuisance parameter and analytically marginalized over in the supernova likelihood.

Each SNe Ia compilation is analyzed separately in combination with the CC and DESI DR2 datasets. Assuming statistical independence among the CC, DESI DR2 and the selected SNe Ia compilation, the total $\chi^2$ for each combined dataset is given by
\begin{equation}\label{eq:3.41}
\chi^2_{\rm tot}=
\chi^2_{\rm CC}
+\chi^2_{\rm DESI\,DR2}
+\chi^2_{SNe},
\end{equation}
corresponding to the CC + DESI~DR2 + Pantheon+, CC + DESI~DR2 + DES-Dovekie and CC + DESI~DR2 + Union3 dataset combinations, respectively.

\begin{figure}
\renewcommand{\figurename}{Fig.}
\centering
\includegraphics[width=0.85\textwidth]{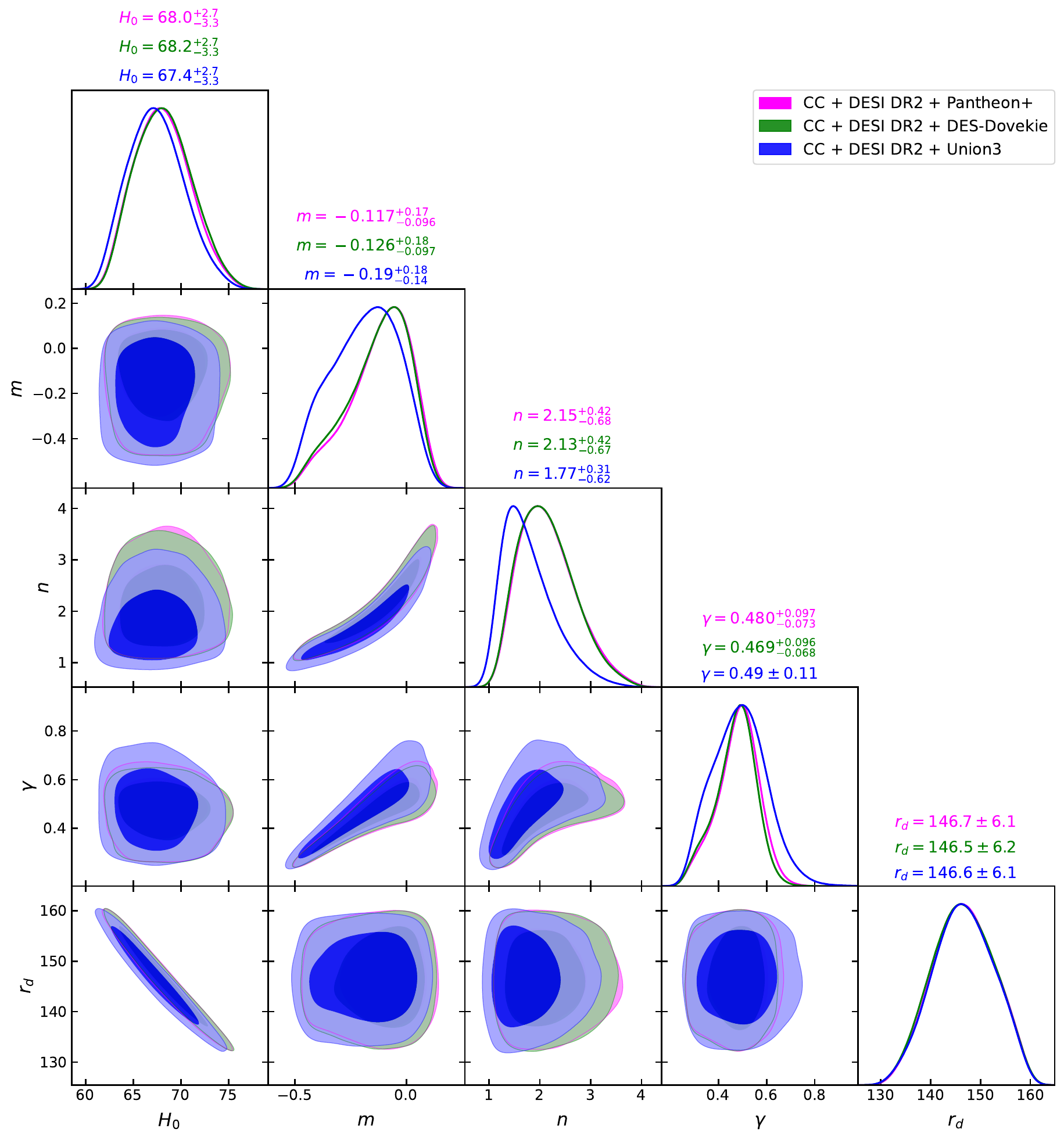}
\caption{\justifying One-dimensional marginalized distributions (diagonal panels) and two-dimensional correlation contours ($68\%$ and $95\%$ credible regions) for the parameter set $(H_0,\,m,\,n,\,\gamma,\,r_d)$, comparing the CC + DESI DR2 constraints combined separately with the Pantheon+, DES-Dovekie and Union3 supernova samples.}
\label{fig:3.01}
\end{figure}

\setlength{\tabcolsep}{3pt}
\begin{table*}[!htbp]
\centering
\renewcommand{\arraystretch}{1.6}
\setlength{\tabcolsep}{5pt}
\small
\caption{\justifying Marginalized posterior constraints for the parameters $(H_0, m, n, \gamma, r_d)$ for the different dataset combinations, together with their $1\sigma$ credible intervals, obtained from the MCMC analysis.}
\label{tab:3.02}
\begin{tabular}{lccccc}
\hline
 Dataset  &  $H_0\,[\mathrm{km\,s^{-1}\,Mpc^{-1}}]$  &  $m$  &  $n$ & $\gamma$ & $r_d\,[\mathrm{Mpc}]$   \\
\hline
 CC + DESI DR2 + Pantheon+ & 
 $68.0^{+2.7}_{-3.3}$ & 
 $-0.117^{+0.17}_{-0.096}$ & 
 $2.15^{+0.42}_{-0.68}$ & 
 $0.480^{+0.097}_{-0.073}$ & 
 $146.7\pm 6.1$ \\
 CC + DESI DR2 + DES-Dovekie  &  
 $68.2^{+2.7}_{-3.3}$   & 
 $-0.126^{+0.18}_{-0.097}$  & 
 $2.13^{+0.42}_{-0.67}$  &  
 $0.469^{+0.096}_{-0.068}$  & 
 $146.5\pm 6.2$  \\
 CC + DESI DR2 + Union3  &  
 $67.4^{+2.7}_{-3.3}$   & 
 $-0.19^{+0.18}_{-0.14}$  & 
 $1.77^{+0.31}_{-0.62}$ & 
 $0.49\pm 0.11$ &  
 $146.6\pm 6.1$  \\
\hline
\end{tabular}
\end{table*}

\subsection{Results and discussion}\label{subsec:3.04}
Table~\ref{tab:3.02} summarizes the marginalized posterior constraints on the cosmological parameters obtained from the joint analysis of the three datasets, while the corresponding 1D and 2D posterior probability distributions are illustrated in Fig.~\ref{fig:3.01}. The posterior contours are overall well localized, although significant correlations among several parameters are evident, indicating parameter degeneracies within the model.

A particularly robust result is obtained for the Hubble constant. The three combinations,
CC + DESI~DR2 + Pantheon+,
CC + DESI~DR2 + DES-Dovekie and
CC + DESI~DR2 + Union3, give
$68.0^{+2.7}_{-3.3}$,
$68.2^{+2.7}_{-3.3}$ and
$67.4^{+2.7}_{-3.3}$
$\mathrm{km\,s^{-1}\,Mpc^{-1}}$, respectively. The substantial overlap of their $1\sigma$ credible intervals indicates that the inferred present-day expansion rate is largely insensitive to the choice of SNe Ia compilation. These values are also consistent with the CMB-based determination from Planck 2018 within the quoted uncertainties~\citep{Aghanim_2020}. The sound horizon scale is also reasonably stable, with
$146.7\pm 6.1 \,\mathrm{Mpc}$, $146.5\pm 6.2 \,\mathrm{Mpc}$ and
$146.6\pm 6.1\,\mathrm{Mpc}$
for CC + DESI~DR2 + Pantheon+, CC + DESI~DR2 + DES-Dovekie and CC + DESI~DR2 + Union3, respectively.

\begin{figure}
\renewcommand{\figurename}{Fig.}
\centering
\includegraphics[width=0.98\textwidth]{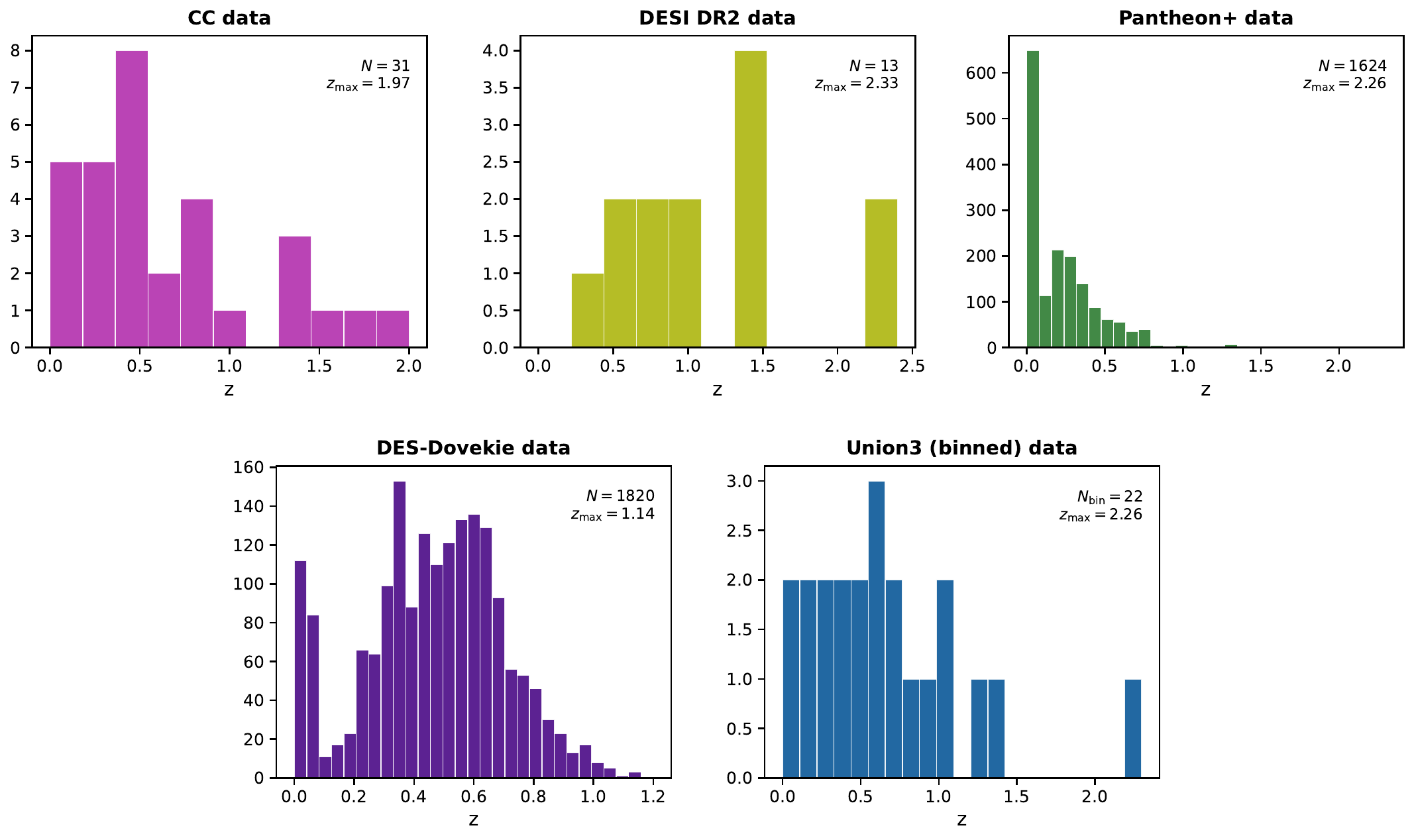}
\caption{\justifying Redshift distributions of the observational datasets employed in the analysis: CC, DESI DR2 BAO, Pantheon+, DES-Dovekie and the binned Union3 compilation. The horizontal axis represents the redshift \(z\), while the vertical axis gives the number of measurements within each redshift interval. The values of \(N\) and \(z_{\rm max}\) indicate the number of observational data points and maximum redshift for each dataset, respectively; for Union3, \(N_{\rm bin}\) denotes the number of adopted redshift bins.}
  \label{fig:3.02}
  \end{figure}

\medskip

\noindent\textbf{Interpretation of the redshift distributions.}
Fig.~\ref{fig:3.02} shows the redshift distributions of the observational datasets used in our analysis. The CC and DESI DR2 samples contain relatively sparse measurements distributed over broad redshift ranges, extending to $z=1.97$ and $z=2.33$, respectively. The Pantheon+ and DES-Dovekie supernova compilations provide substantially denser sampling at low and intermediate redshifts, extending to $z=2.26$ and $z=1.14$, respectively. The Union3 compilation is represented by 22 binned measurements extending to $z=2.26$. Overall, the combined datasets provide broad coverage of the low- and intermediate-redshift universe, supplemented by a smaller number of measurements at higher redshifts. This redshift coverage provides useful leverage for constraining the evolution of the late-time expansion history.

\begin{figure}
\renewcommand{\figurename}{Fig.}
\centering
\includegraphics[width=0.44\textwidth]{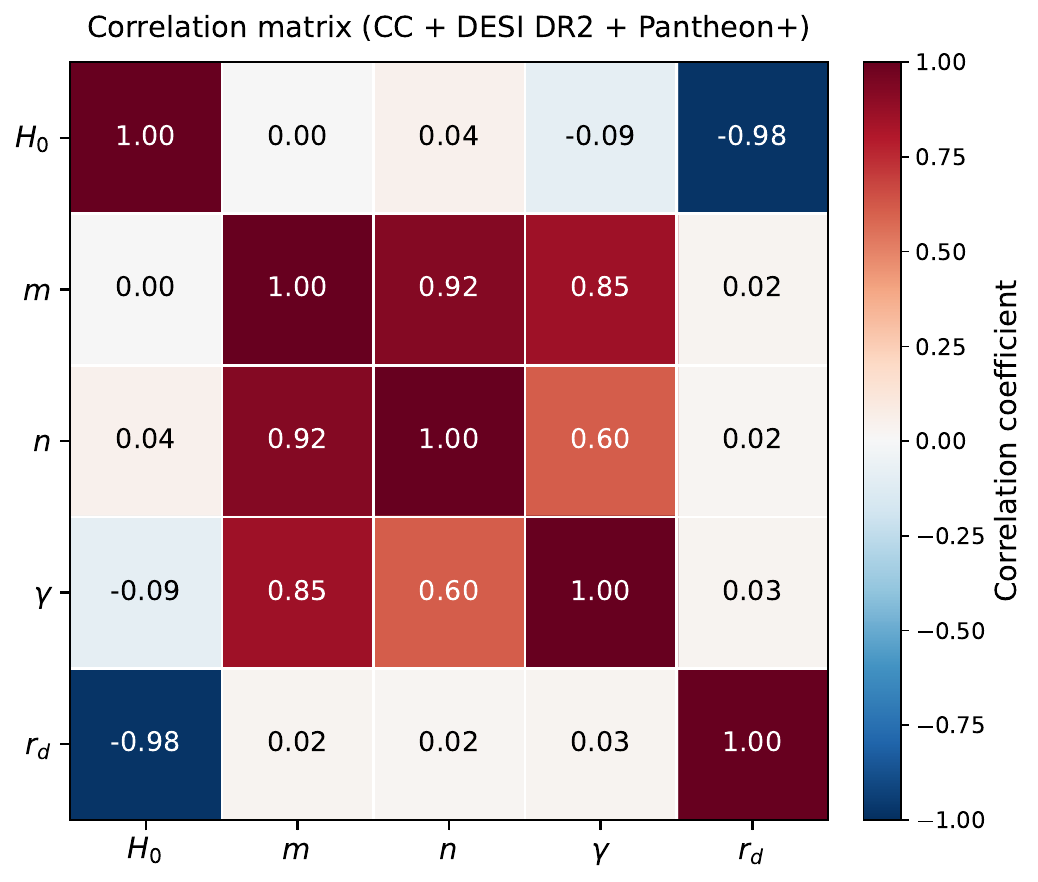}
\hfill
\includegraphics[width=0.44\textwidth]{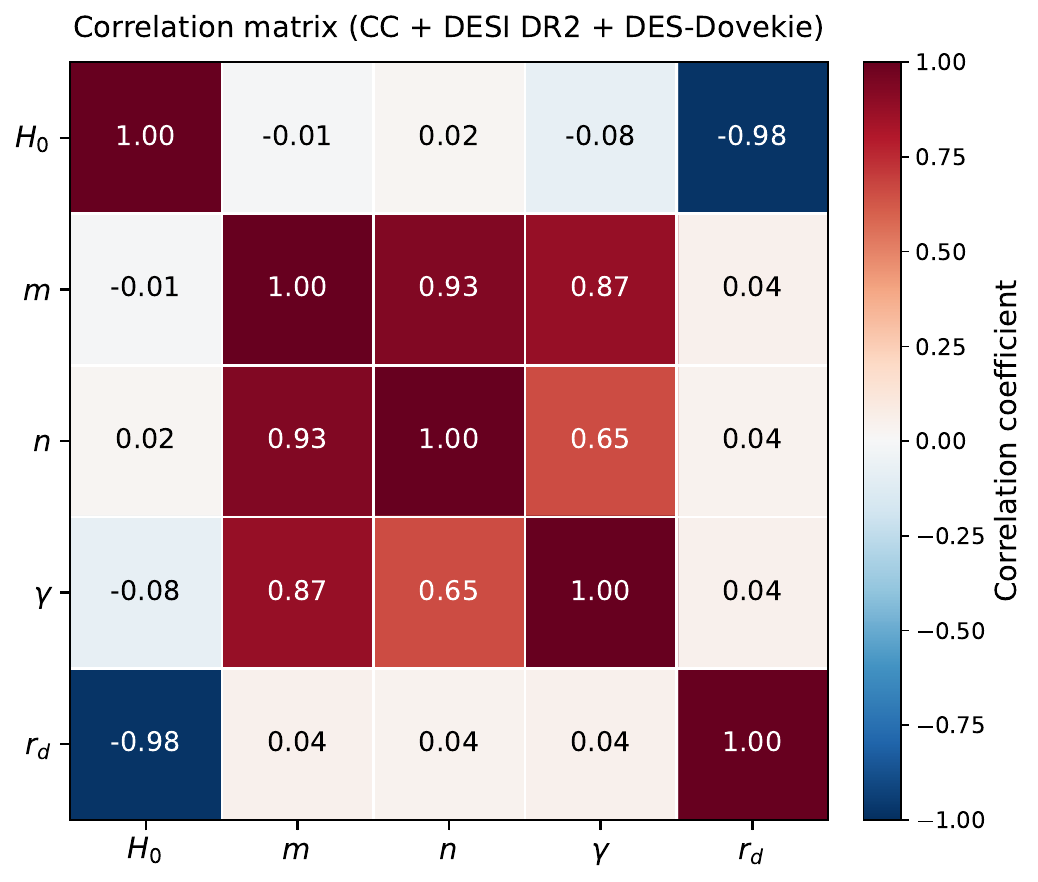}

\vspace{0.1cm}

\makebox[\textwidth][c]{%
\includegraphics[width=0.44\textwidth]{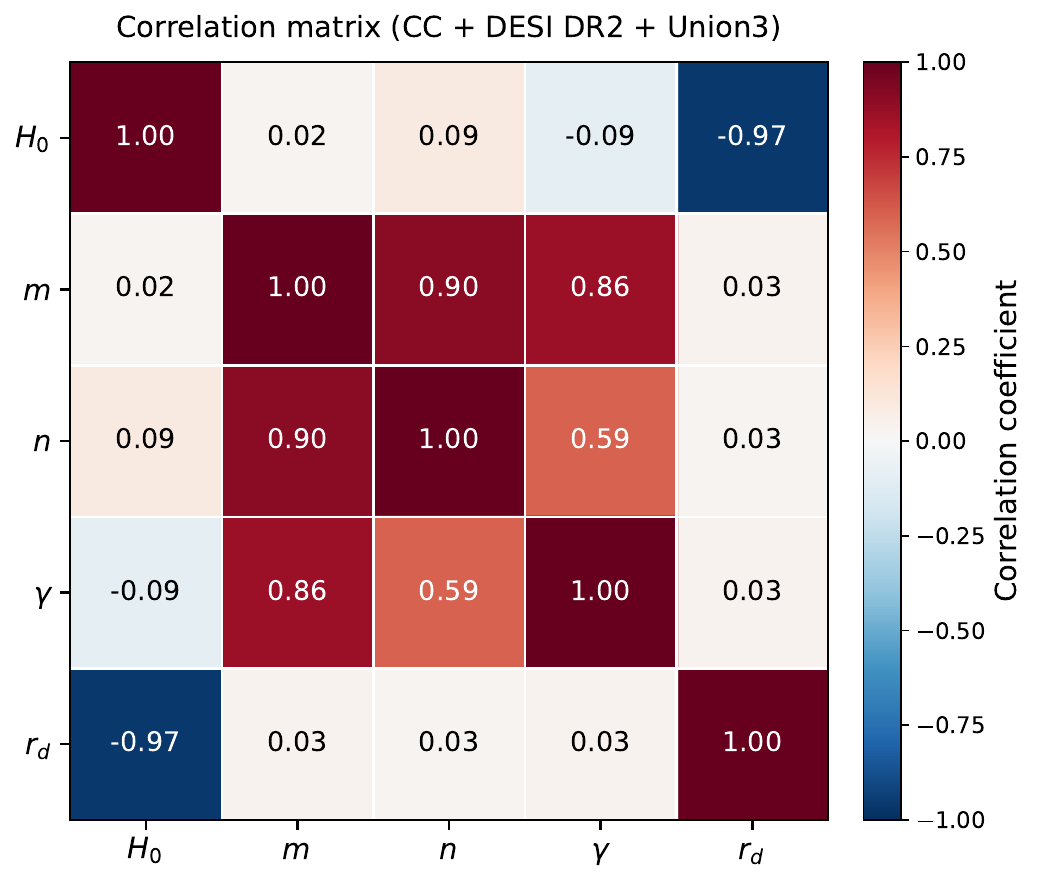}
}
\caption{\justifying Correlation matrices of the model parameters $H_0$, $m$, $n$, $\gamma$ and $r_d$ for the three combined observational datasets: CC + DESI DR2 + Pantheon+, CC + DESI DR2 + DES-Dovekie and CC + DESI DR2 + Union3.}
\label{fig:3.03}
\end{figure}

\medskip
\noindent\textbf{Interpretation of the correlation matrices.}
Fig.~\ref{fig:3.03} shows the correlation matrices for the parameters $(H_0,\,m,\,n,\,\gamma,\,r_d)$ obtained from the three combined datasets, namely CC + DESI DR2 + Pantheon+, CC + DESI DR2 + DES-Dovekie and CC + DESI DR2 + Union3. The color scale represents the strength and sign of the parameter correlations, with positive and negative values
corresponding to correlated and anticorrelated parameter variations, respectively.

A pronounced negative correlation is observed between $H_0$ and $r_d$ across all three dataset combinations, with correlation coefficients ranging from approximately $-0.97$ to $-0.98$. This strong anticorrelation reflects a well-known degeneracy in BAO constraints between the present-day expansion rate ($H_0$) and the sound horizon scale ($r_d$).
Because of this degeneracy, a change in one parameter can be compensated for by a change in the other, leaving the measured distance ratios largely unaffected.

\begin{figure}
\renewcommand{\figurename}{Fig.}
\centering
\includegraphics[width=0.72\textwidth]{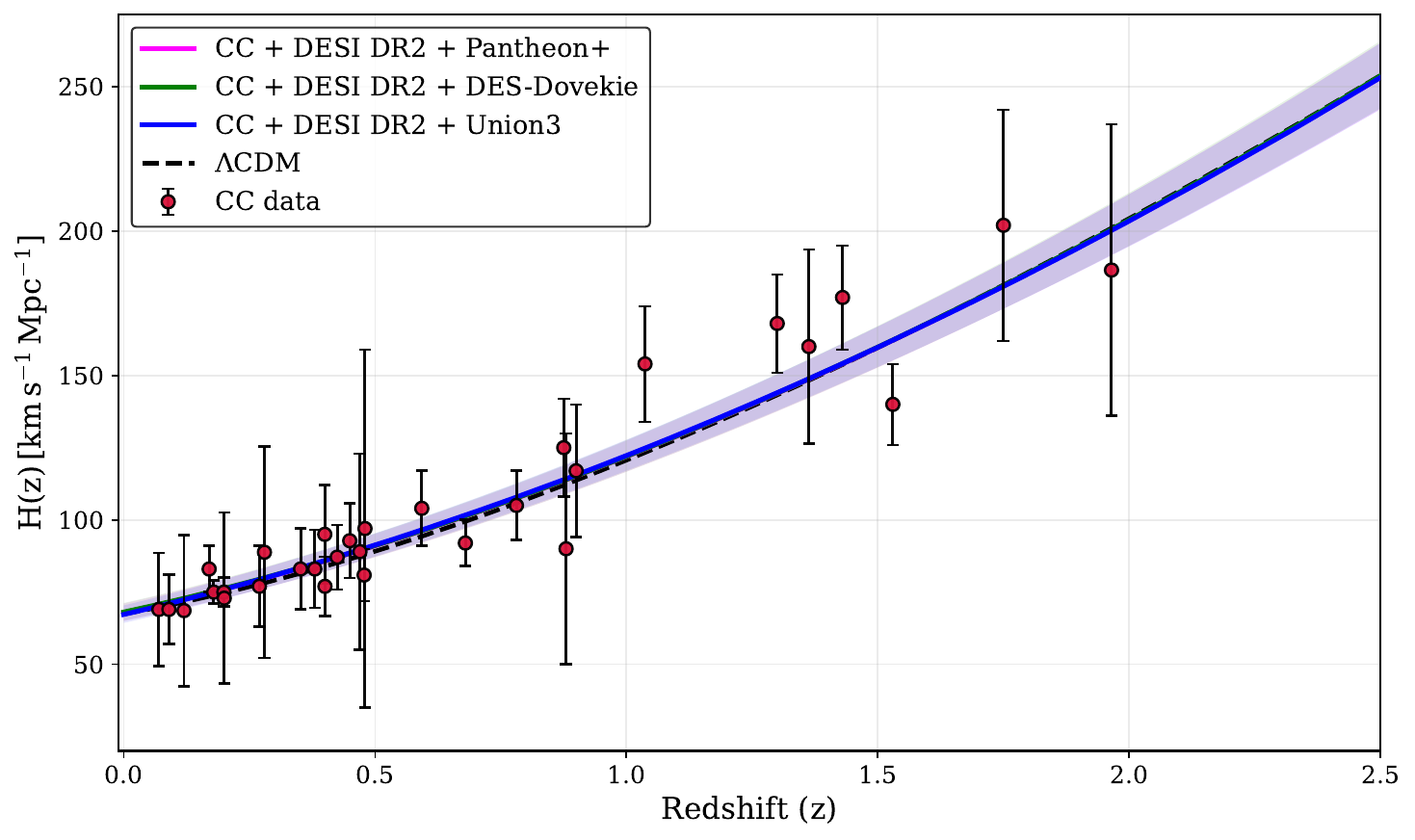}
\caption{\justifying Evolution of the reconstructed Hubble parameter for the CC + DESI DR2 constraints combined separately with the Pantheon+, DES-Dovekie and Union3 supernova samples, shown by the magenta, green and blue solid curves, respectively. The black dashed curve represents $\Lambda$CDM, while the data points with error bars denote the 31 CC measurements. The shaded bands indicate the corresponding $1\sigma$ credible intervals.}
  \label{fig:3.04}
  \end{figure}

\begin{figure}
\renewcommand{\figurename}{Fig.}
\centering
\includegraphics[width=0.72\textwidth]{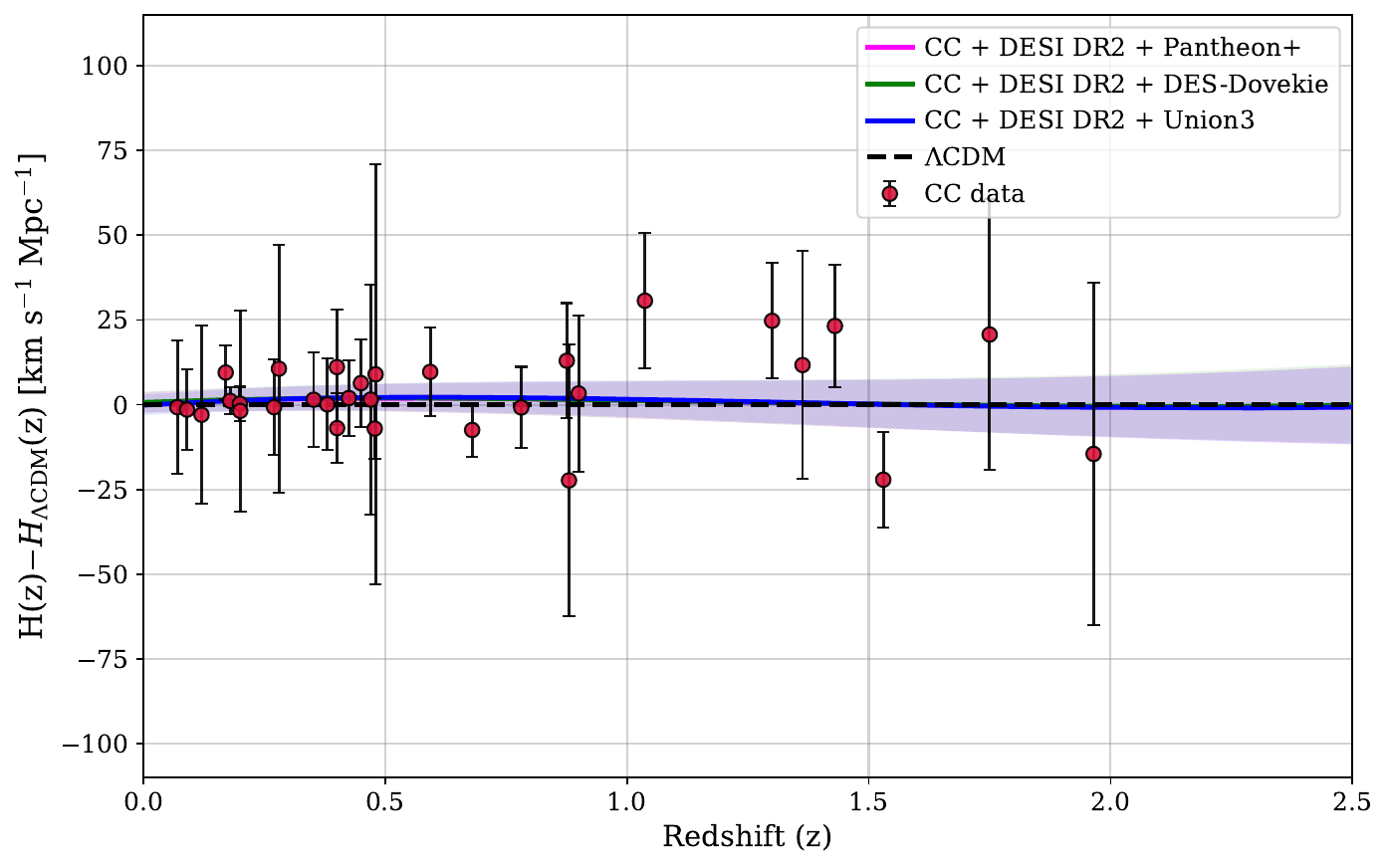}
  \caption{\justifying Difference between the reconstructed Hubble parameter and the $\Lambda$CDM prediction, $H(z)-H_{\Lambda\mathrm{CDM}}(z)$, for the CC + DESI DR2 constraints combined separately with the Pantheon+, DES-Dovekie and Union3 supernova samples, shown by the magenta, green and blue solid curves, respectively. The black solid curve represents the $\Lambda$CDM reference, while the data points with error bars denote the 31 CC measurements.}
  \label{fig:3.05}
  \end{figure}

The parameters $m$, $n$ and $\gamma$ exhibit strong positive correlations. The $m$--$n$ correlation remains consistently strong across the three dataset combinations, with coefficients of approximately $0.92$, $0.93$ and $0.90$, respectively. A similarly strong positive correlation is observed between $m$ and $\gamma$, with coefficients ranging from $0.85$ to $0.87$, while the $n$--$\gamma$ correlation is also positive. These correlations suggest parameter degeneracies within the adopted expansion parametrization, indicating that the parameters $m$, $n$ and $\gamma$ are not independently constrained by the data and that variations in one parameter may be partially compensated by changes in the others. In contrast, both $H_0$ and $r_d$ are nearly uncorrelated with the three expansion-history parameters, $m$, $n$ and $\gamma$. Overall, the correlation structure remains qualitatively consistent across all dataset combinations.

\medskip
\noindent\textbf{Impact of combined observational data on the reconstructed $H(z)$ and comparison with $\Lambda$CDM.}
Fig.~\ref{fig:3.04} shows the evolution of the reconstructed Hubble parameter $H(z)$ for the CC + DESI DR2 + Pantheon+, CC + DESI DR2 + DES-Dovekie and CC + DESI DR2 + Union3 dataset combinations. The three reconstructions closely follow the CC measurements and exhibit similar expansion histories, remaining broadly consistent with the $\Lambda$CDM prediction. The corresponding $1\sigma$ credible bands largely overlap across the three dataset combinations, suggesting that the reconstructed expansion histories are consistent within their posterior uncertainties over most of the redshift range. These shaded regions represent the posterior uncertainties in $H(z)$ and illustrate the constraints provided by the combined observational datasets. Overall, the results indicate that the three dataset combinations yield similar reconstructions of the cosmic expansion history without substantial differences within the displayed uncertainties.

Fig.~\ref{fig:3.05} presents the relative differences between the reconstructed Hubble parameter and the corresponding $\Lambda$CDM prediction for the three dataset combinations. The deviations remain generally close to zero over the redshift range considered, indicating that the reconstructed expansion histories are broadly consistent with $\Lambda$CDM. This agreement is particularly evident at low redshifts, where the differences are small. Overall, the relative-difference analysis indicates that all three dataset combinations yield only mild departures from the $\Lambda$CDM prediction.

\begin{figure}
\renewcommand{\figurename}{Fig.}
\centering
\includegraphics[width=0.48\textwidth]{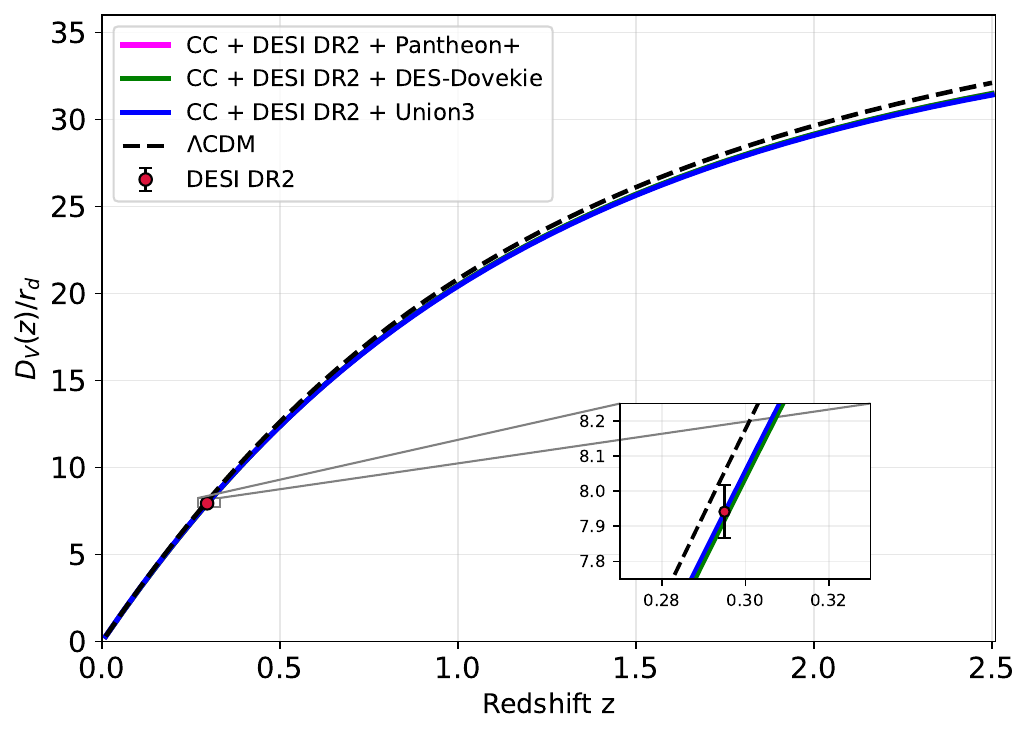}
\hfill
\includegraphics[width=0.48\textwidth]{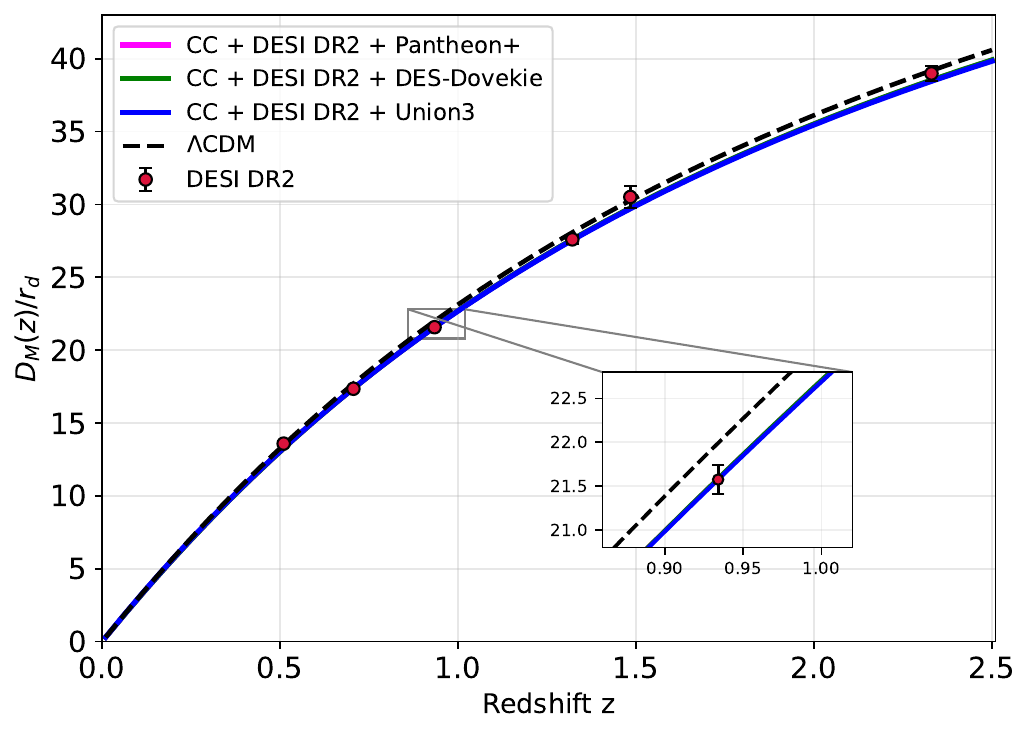}

\vspace{0.3cm}

\makebox[\textwidth][c]{%
\includegraphics[width=0.58\textwidth]{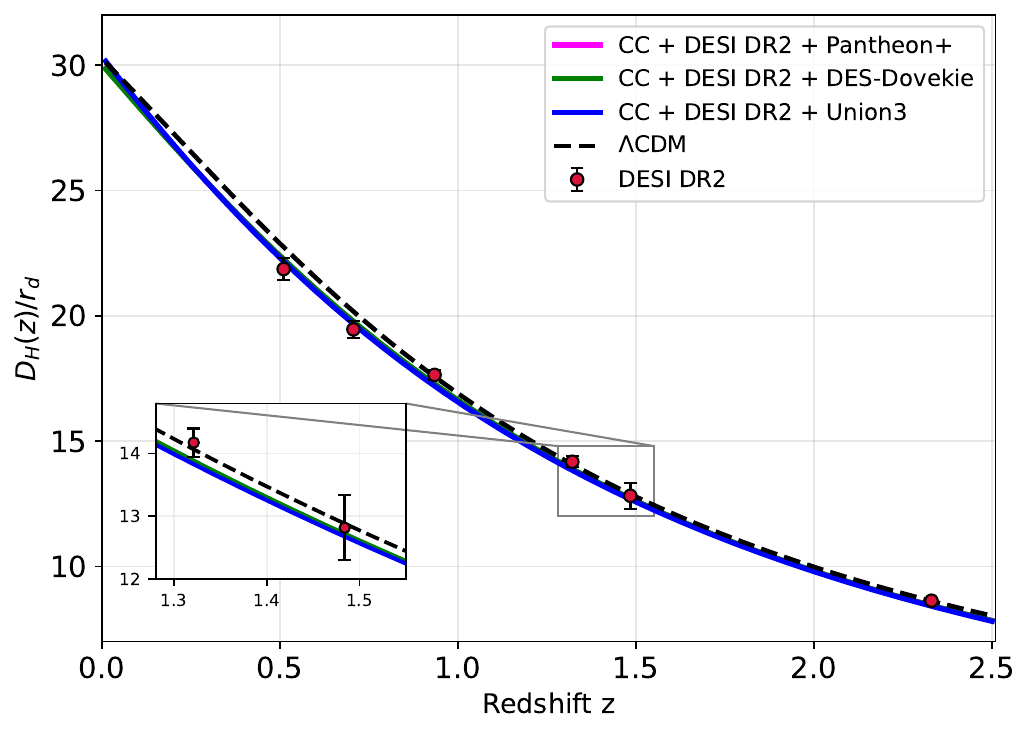}
}
\caption{\justifying Evolution of the reconstructed BAO distance ratios $D_{\rm V}(z)/r_d$, $D_{\rm M}(z)/r_d$ and $D_{\rm H}(z)/r_d$, shown in the top-left, top-right and bottom panels, respectively, for the CC + DESI DR2 constraints combined separately with the Pantheon+, DES-Dovekie and Union3 supernova samples. The magenta, green and blue solid curves denote the corresponding reconstructions, while the black dashed curve represents $\Lambda$CDM. The data points with error bars denote the DESI DR2 BAO measurements, while the insets highlight regions where the reconstructed curves differ only slightly.}
\label{fig:3.06}
\end{figure}

Fig.~\ref{fig:3.06} presents the reconstructed BAO distance ratios, $D_{\rm V}(z)/r_d$, $D_{\rm M}(z)/r_d$ and $D_{\rm H}(z)/r_d$, for the three dataset combinations. The reconstructions are compared with the $\Lambda$CDM prediction and the DESI DR2 BAO measurements. All three reconstructed distance ratios closely reproduce the DESI DR2 data and remain broadly consistent with $\Lambda$CDM across the redshift range. The reconstructions are nearly indistinguishable from one another, with only small dataset-dependent differences visible in the magnified insets. This indicates that the inferred BAO distance scales are largely insensitive to the choice of supernova compilation and are robust across the three dataset combinations.

\subsection{Statistical model comparison}\label{subsec:3.05}
To assess the relative statistical performance of the proposed $f(Q)$ model and the standard $\Lambda$CDM scenario, we employ model selection techniques that balance the goodness of fit against model complexity. While the minimum chi-square statistic, $\chi^2_{\min}$, quantifies the agreement between the theoretical predictions and observational data, it does not penalize models with extra free parameters. To account for this, we evaluate both the Akaike Information Criterion (AIC)~\citep{Akaike_1974} and the Bayesian Information Criterion (BIC)~\citep{Schwarz_1978}. The AIC and BIC for a model with $k$ free parameters constrained by $N$ observational data points are defined as~\citep{Rezaei_2021}
\begin{equation}\label{eq:3.42}
\mathrm{AIC} = \chi^2_{\min} + 2k, \qquad
\mathrm{BIC} = \chi^2_{\min} + k\ln N.
\end{equation}
The AIC imposes a constant penalty of $2$ for each additional free parameter, whereas the BIC penalty increases with the dataset size through the $\ln N$ term. Consequently, for $N>e^2\simeq 7.39$, or equivalently for integer $N\geq 8$, the BIC imposes a larger penalty per additional free parameter than the AIC.

The relative performance of the two cosmological scenarios can be assessed through the differences in their information criteria,
\begin{equation}\label{eq:3.43}
\Delta\mathrm{AIC} =
\mathrm{AIC}_{\mathrm{model}}-\mathrm{AIC}_{\Lambda\mathrm{CDM}},
\qquad
\Delta\mathrm{BIC} =
\mathrm{BIC}_{\mathrm{model}}-\mathrm{BIC}_{\Lambda\mathrm{CDM}}.
\end{equation}
A negative value of either quantity favors the proposed $f(Q)$ model, while a positive value favors $\Lambda$CDM. The magnitude of the difference indicates the degree to which the two models are statistically distinguishable. Following the commonly adopted criteria, $|\Delta\mathrm{AIC}|\leq2$ indicates that the two models have comparable support, whereas $4\leq|\Delta\mathrm{AIC}|\leq7$ suggests considerably less support for the model with the larger AIC. A value of $|\Delta\mathrm{AIC}|>10$ indicates essentially no support for the model having the larger AIC~\citep{Burnham_2004}. For the BIC, the ranges $0\leq|\Delta\mathrm{BIC}|<2$, $2\leq|\Delta\mathrm{BIC}|<6$, $6\leq|\Delta\mathrm{BIC}|<10$ and $|\Delta\mathrm{BIC}|\geq10$ correspond, respectively, to weak, positive, strong and very strong evidence against the model with the larger BIC~\citep{Kass_1995}. Hence, for each dataset, the model with the smaller AIC or BIC value is statistically favored.

\begin{table}[ht]
\centering
\renewcommand{\arraystretch}{1.8}
\setlength{\tabcolsep}{9pt} 
\small
\caption{\justifying Statistical model comparison between the proposed $f(Q)$ model and the flat $\Lambda$CDM model for the CC + DESI DR2 + Pantheon+, CC + DESI DR2 + DES-Dovekie and CC + DESI DR2 + Union3 dataset combinations.}
\label{tab:3.03}
\begin{tabular}{llccccc}
\hline
Dataset & Model & $\chi^2_{\min}$  & AIC & BIC & $\Delta$AIC & $\Delta$BIC \\
\hline

\multirow{2}{*}{\makecell[l]{CC + DESI DR2 + Pantheon+\\$(N=1668)$}}
& $f(Q)$ model
& 1479.23
& 1489.23
& 1516.32
& $-1.84$
& $+8.99$ \\

& $\Lambda$CDM
& 1485.07
& 1491.07
& 1507.33
& 0
& 0 \\

\hline

\multirow{2}{*}{\makecell[l]{CC + DESI DR2 + DES-Dovekie\\$(N=1864)$}}
& $f(Q)$ model
& 1653.67
& 1663.67
& 1691.33
& $-2.18$
& $+8.89$ \\

& $\Lambda$CDM
& 1659.85
& 1665.85
& 1682.44
& 0
& 0 \\

\hline

\multirow{2}{*}{\makecell[l]{CC + DESI DR2 + Union3\\$(N=66)$}}
& $f(Q)$ model
& 44.10
& 54.09
& 65.05
& $-5.29$
& $-0.90$ \\

& $\Lambda$CDM
& 53.38
& 59.38
& 65.95
& 0
& 0 \\

\hline
\end{tabular}
\end{table}

The statistical model comparison presented in Table~\ref{tab:3.03} shows that, for all three dataset combinations, the proposed model yields a lower minimum chi-square than the flat $\Lambda$CDM model, indicating a better best-fit description of the observational data. The AIC values are also lower for the proposed model, with $\Delta\mathrm{AIC}=-1.84$, $-2.18$ and $-5.29$ for the CC + DESI DR2 + Pantheon+, CC + DESI DR2 + DES-Dovekie and CC + DESI DR2 + Union3 combinations, respectively. These results indicate a preference for the proposed model according to the AIC, although the strength of this preference varies across the datasets. In contrast, the BIC yields $\Delta\mathrm{BIC}=+8.99$, $+8.89$ and $-0.90$, respectively. The positive values of $\Delta\mathrm{BIC}$ for the CC + DESI DR2 + Pantheon+ and CC + DESI DR2 + DES-Dovekie combinations favor the flat $\Lambda$CDM model, whereas the negative value of $\Delta\mathrm{BIC}$ for the CC + DESI DR2 + Union3 combination indicates only a negligible preference for the proposed model. This difference between the AIC and BIC reflects their different penalties for model complexity, with the BIC imposing a stronger penalty on additional free parameters for sufficiently large datasets. Overall, although the proposed model provides a better best-fit description of the data and is favored by the AIC for all three dataset combinations, the BIC favors flat $\Lambda$CDM for the Pantheon+ and DES-Dovekie combinations and shows essentially no preference between the models for Union3 combination.

\section{Cosmographic parameters}\label{sec:3.05}
To investigate the late-time cosmic expansion and its evolution, we employ cosmography. Cosmography provides a geometrical description of cosmic expansion in terms of successive time derivatives of the scale factor, without requiring a specific dark energy model \citep{Visser_2004}. It therefore provides a useful complementary diagnostic for characterizing the reconstructed expansion history and comparing it with the standard $\Lambda$CDM scenario.

The deceleration, jerk and snap parameters are defined as
\begin{equation}\label{eq:3.44}
q=-\frac{\ddot{a}}{aH^2}, \qquad
j=\frac{a^{(3)}}{aH^3}, \qquad
s_p=\frac{a^{(4)}}{aH^4},
\end{equation}
respectively. Thus, $q$, $j$ and $s_p$ characterize the second-, third- and
fourth-order kinematical properties of the cosmic expansion
\citep{Visser_2004}. Using
$\frac{d}{dt}=-(1+z)H\frac{d}{dz},$
these parameters can be written in terms of the redshift dependence of the Hubble parameter as

\begin{equation}\label{eq:3.45}
q(z) = (1+z)\frac{1}{H}\frac{dH}{dz}-1,
\end{equation}

\begin{equation}\label{eq:3.46}
j(z) = (1 + z)\frac{dq}{dz} + q\left(1 + 2q\right),
\end{equation}
and
\begin{equation}\label{eq:3.47}
s_p(z) = -(1 + z)\frac{dj}{dz}
-j\left(2 + 3q\right).
\end{equation}

These parameters provide a hierarchical description of the cosmic expansion history. The deceleration parameter determines whether the universe is accelerating or decelerating, while the jerk and snap parameters characterize successively higher-order derivatives of the expansion. In particular, the jerk parameter serves as a useful diagnostic of deviations from the standard $\Lambda$CDM expansion history, for which $j_{\Lambda{\rm CDM}}=1$. The snap parameter provides an additional fourth-order diagnostic, allowing higher-order features of the expansion history to be examined that are not captured by the deceleration and jerk parameters alone \citep{Visser_2004}.

\begin{figure}
\renewcommand{\figurename}{Fig.}
\centering
\includegraphics[width=0.74\textwidth]{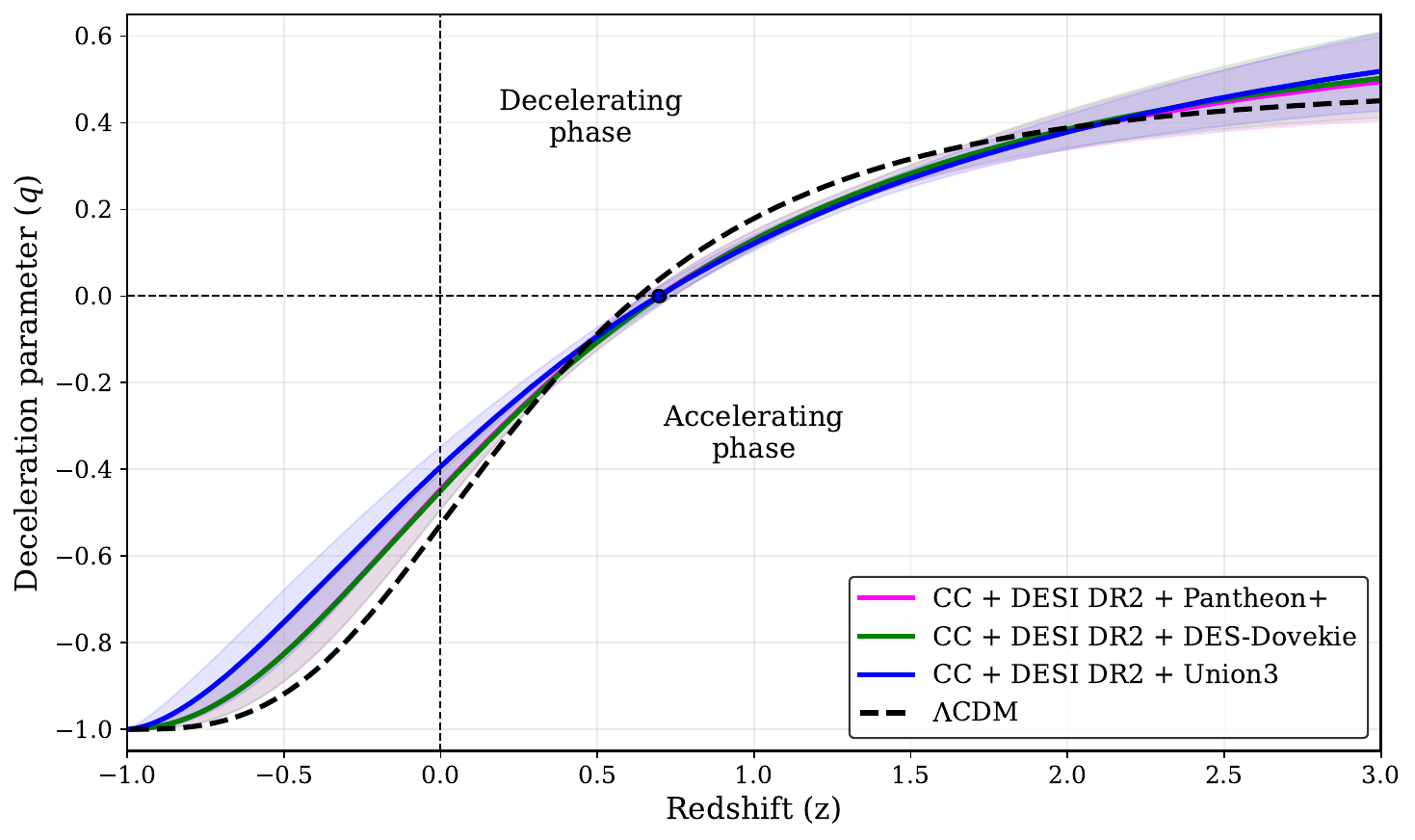}
\caption{\justifying Deceleration parameter as a function of redshift $z$ for the $f(Q)$ model compared to the standard $\Lambda\text{CDM}$ cosmology, constrained by the CC + DESI DR2 data combined separately with the Pantheon+, DES-Dovekie and Union3 Type Ia supernova datasets.}
  \label{fig:3.07}
  \end{figure}

\begin{figure}
\renewcommand{\figurename}{Fig.}
\centering
\includegraphics[width=0.74\textwidth]{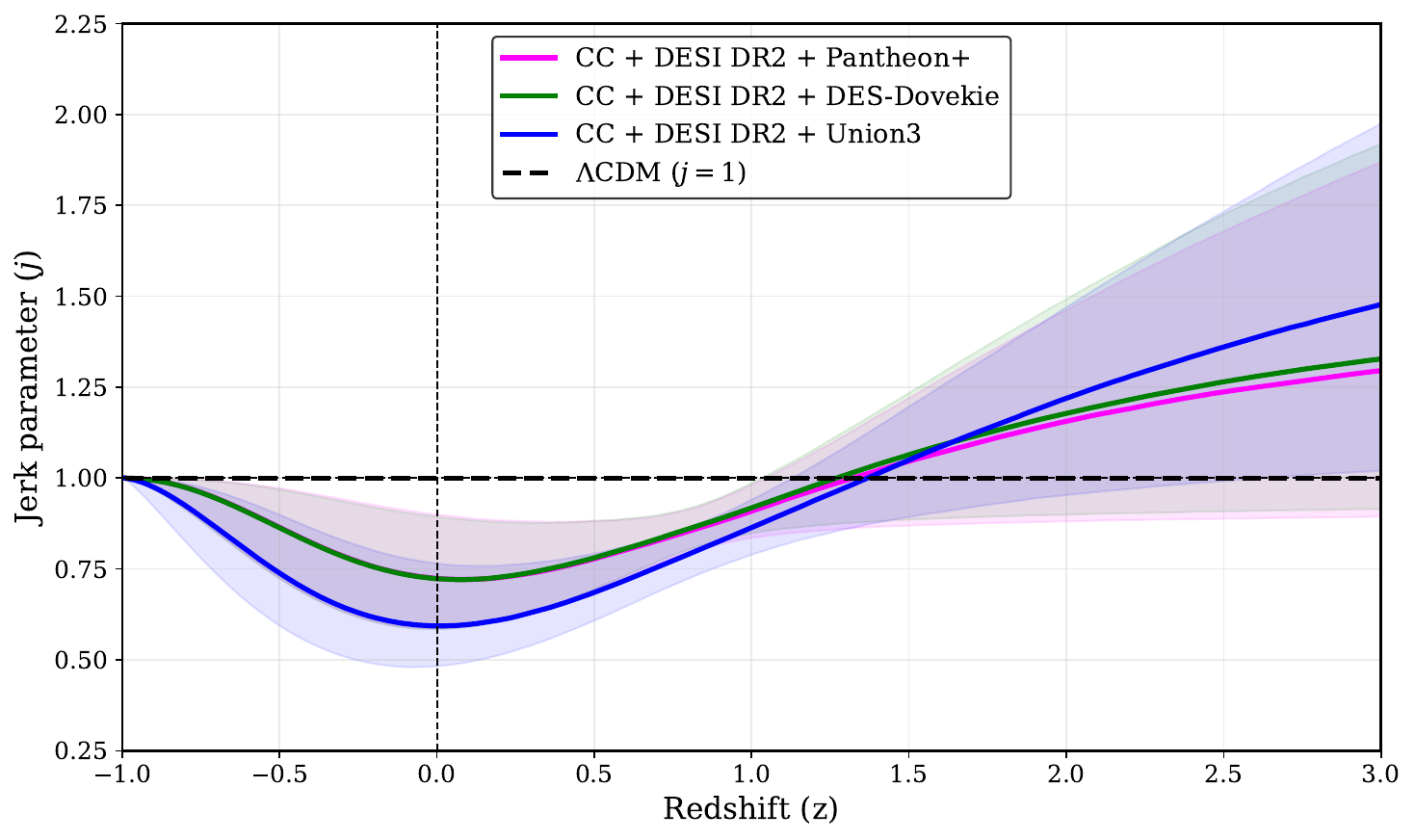}
  \caption{\justifying Jerk parameter as a function of redshift $z$ for the $f(Q)$ model compared to the standard $\Lambda\text{CDM}$ cosmology, constrained by the CC + DESI DR2 data combined separately with the Pantheon+, DES-Dovekie and Union3 Type Ia supernova datasets.}
  \label{fig:3.08}
  \end{figure}

Fig.~\ref{fig:3.07} presents the corresponding evolution of the deceleration parameter $q(z)$. All three cases exhibit a transition from a decelerating phase at higher redshifts, characterized by $q(z)>0$, to accelerated expansion at late times, where $q(z)<0$. The present-day values are $q_0=-0.446^{+0.040}_{-0.047}$, $-0.449^{+0.040}_{-0.045}$ and $-0.394^{+0.046}_{-0.055}$, respectively, confirming accelerated expansion at the current epoch and remaining consistent with recent observational results~\citep{Bhagat_2025,Mishra_2025,Pourojaghi_2025}. The corresponding transition redshifts are $z_t=0.696^{+0.046}_{-0.048}$, $0.700^{+0.044}_{-0.048}$ and $0.695^{+0.041}_{-0.045}$, respectively, placing the deceleration-to-acceleration transition within the relatively narrow range $0.695\leq z_t\leq0.700$~\citep{Singh_2027,Mishra_2025}. The close agreement among the three values of $z_t$ indicates that the inferred transition to late-time acceleration is only weakly dependent on the choice of supernova compilation.

The evolution of jerk parameter $j(z)$ is shown in Fig.~\ref{fig:3.08}. Since flat $\Lambda$CDM predicts $j_{\Lambda{\rm CDM}}=1$, it provides a useful higher-order kinematical diagnostic for comparing the reconstructed expansion history with the standard $\Lambda$CDM scenario. The present-day values are $j_0=0.724^{+0.176}_{-0.138}$, $0.723^{+0.171}_{-0.139}$ and $0.593^{+0.172}_{-0.111}$, respectively, for the three dataset combinations. Thus, all three cases yield central values of $j_0$ below the $\Lambda$CDM expectation.

\begin{figure}
\renewcommand{\figurename}{Fig.}
\centering
\includegraphics[width=0.74\textwidth]{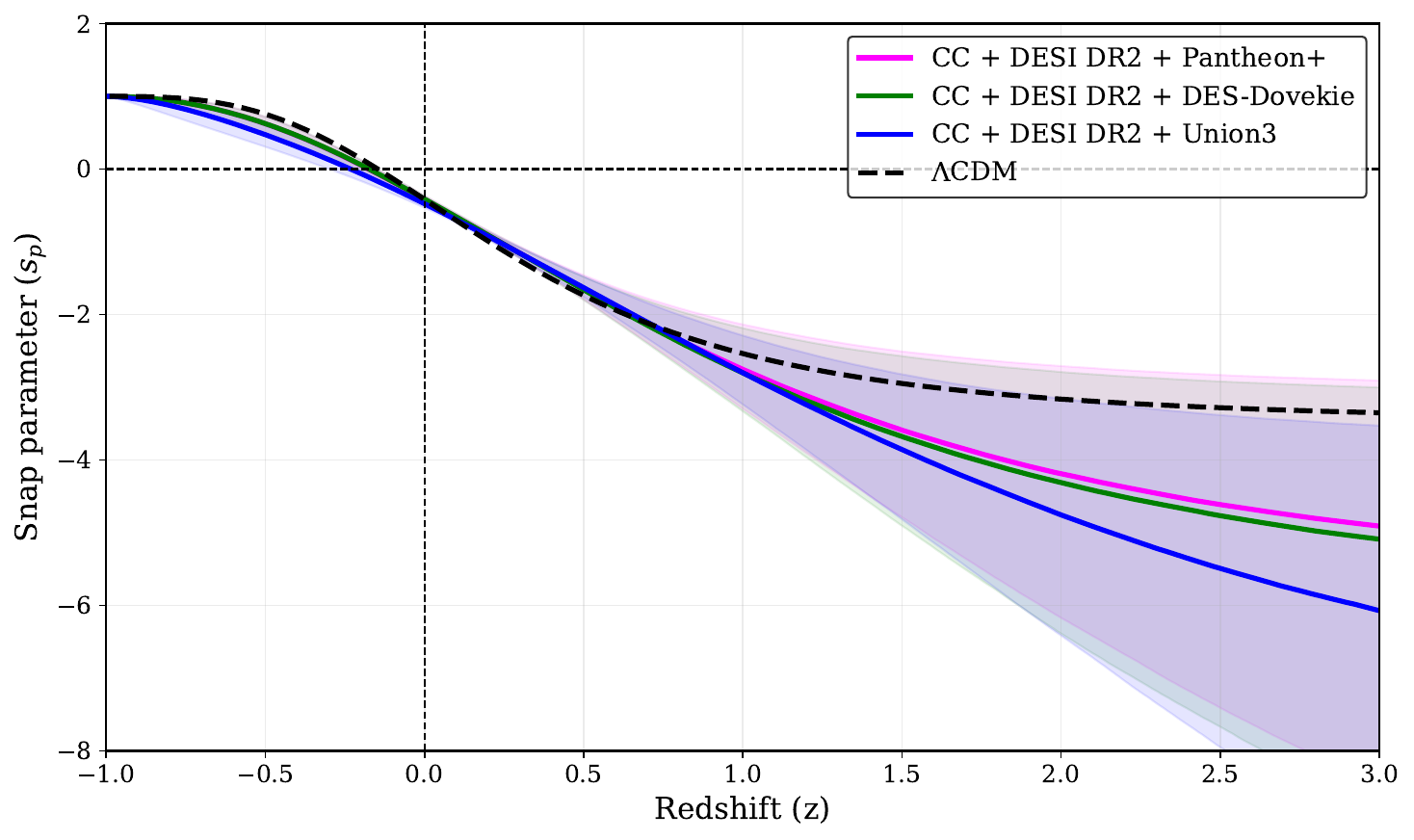}
  \caption{\justifying Snap parameter as a function of redshift $z$ for the $f(Q)$ model compared to the standard $\Lambda\text{CDM}$ cosmology, constrained by the CC + DESI DR2 data combined separately with the Pantheon+, DES-Dovekie and Union3 Type Ia supernova datasets.}
  \label{fig:3.09}
  \end{figure}

The snap parameter $s_p(z)$ is shown in Fig.~\ref{fig:3.09}. The present-day values are $s_p(z=0)$= $-0.415^{+0.048}_{-0.053}$, $-0.412^{+0.046}_{-0.052}$ and $-0.477^{+0.069}_{-0.055}$, respectively, remain relatively close across the three dataset combinations. Unlike the jerk parameter, the $\Lambda$CDM snap parameter is redshift-dependent and approaches the de Sitter limit, $s_p\to 1$, as $z\to -1$. The three reconstructed curves likewise approach this limit toward the far future, while their differences become more pronounced at higher redshifts. The CC + DESI DR2 + Pantheon+ and CC + DESI DR2 + DES-Dovekie reconstructions exhibit comparatively moderate declines, whereas the CC + DESI DR2 + Union3 reconstruction shows a substantially steeper decline toward negative values for $z\gtrsim1$. This greater separation among the reconstructed $s_p(z)$ curves highlights the sensitivity of the snap parameter to differences in the reconstructed expansion history, relative to the lower-order cosmographic parameters. Table~\ref{tab:3.04} summarizes the present-day values of the cosmographic parameters and transition redshifts for the three combined datasets.

\begin{table}[ht]
\centering
\renewcommand{\arraystretch}{1.5}
\setlength{\tabcolsep}{12pt}
\caption{\justifying Present-day values of the cosmographic parameters and the transition redshift obtained from the CC + DESI DR2 constraints combined separately with the Pantheon+, DES-Dovekie and Union3 supernova compilations.}
\label{tab:3.04}
\begin{tabular}{lcccc}
\hline
Dataset & \multicolumn{3}{c}{Cosmographic parameters} & \multirow{2}{*}{$z_t$} \\
\cline{2-4}
 & $q_0$ & $j_0$ & $s_0$ & \\
\hline
CC + DESI DR2 + Pantheon+
& $-0.446^{+0.040}_{-0.047}$ 
& $0.724^{+0.176}_{-0.138}$ 
& $-0.415^{+0.048}_{-0.053}$ 
& $0.696^{+0.046}_{-0.048}$ \\
CC + DESI DR2 + DES-Dovekie
& $-0.449^{+0.040}_{-0.045}$ 
& $0.723^{+0.171}_{-0.139}$ 
& $-0.412^{+0.046}_{-0.052}$ 
& $0.700^{+0.044}_{-0.048}$ \\
CC + DESI DR2 + Union3
& $-0.394^{+0.046}_{-0.055}$ 
& $0.593^{+0.172}_{-0.111}$ 
& $-0.477^{+0.069}_{-0.055}$ 
& $0.695^{+0.041}_{-0.045}$ \\
\hline
\end{tabular}
\end{table}

Overall, the cosmographic analysis shows that the inferred late-time acceleration and its transition redshift are relatively robust against the choice of supernova compilation, as indicated by the consistent $q_0$ and $z_t$ values. In contrast, the higher-order parameters $j(z)$ and $s_p(z)$ exhibit increasingly pronounced dependence on the dataset combination, particularly at higher redshifts, highlighting their sensitivity to differences in the reconstructed expansion history. These results provide useful kinematical diagnostics for characterizing the reconstructed $f(Q)$ expansion history without requiring a specific dark energy model.

\section{Observationally constrained evolution of the effective energy density, pressure and equation of state parameter}\label{sec:3.06}
For the proposed model, the effective energy density and cosmic pressure, expressed as functions of redshift, are obtained as
\begin{equation}\label{eq:3.48}
    \rho(z)= 2^m 3^{m+1} (2 m+1)\beta \Bigg[H_0^2 \left(\frac{\gamma  (z+1)^n+1}{\gamma +1}\right)^{\frac{3}{m n+n}}\Bigg]^{m+1},
\end{equation}
and
\begin{equation}\label{eq:3.49}
    p(z)= -\frac{2^m 3^{m+1} (2 m+1)\beta \Bigg[H_0^2 \left(\frac{\gamma  (z+1)^n+1}{\gamma +1}\right)^{\frac{3}{m n+n}}\Bigg]^{m+1}}{\gamma  (z+1)^n+1}
\end{equation}

\begin{figure}
\renewcommand{\figurename}{Fig.}
\centering
\includegraphics[width=0.74\textwidth]{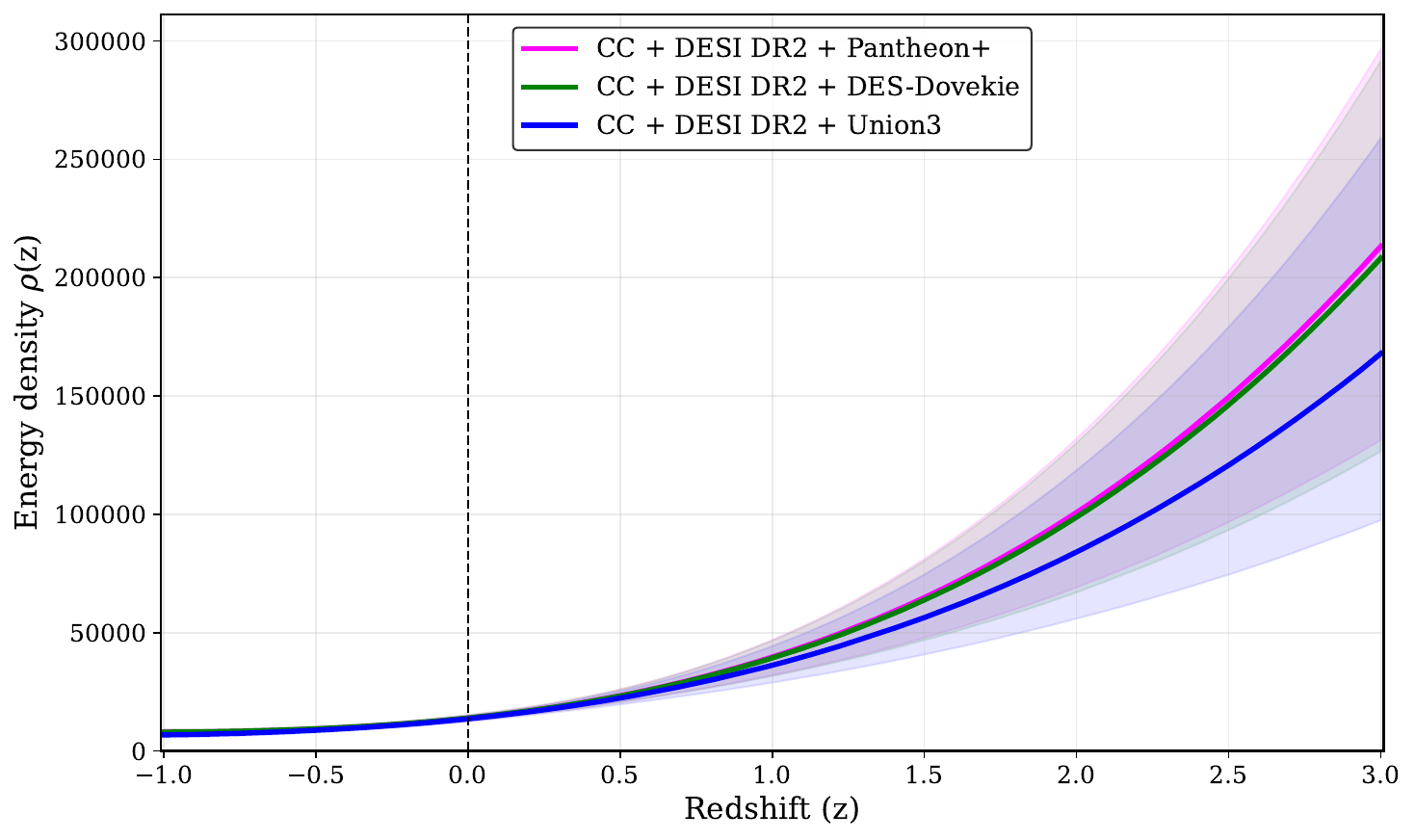}
  \caption{\justifying Evolution of the effective energy density constrained by the CC + DESI DR2 data combined separately with the Pantheon+, DES-Dovekie and Union3 Type Ia supernova datasets.}
  \label{fig:3.10}
  \end{figure}

\begin{figure}
\renewcommand{\figurename}{Fig.}
\centering
\includegraphics[width=0.74\textwidth]{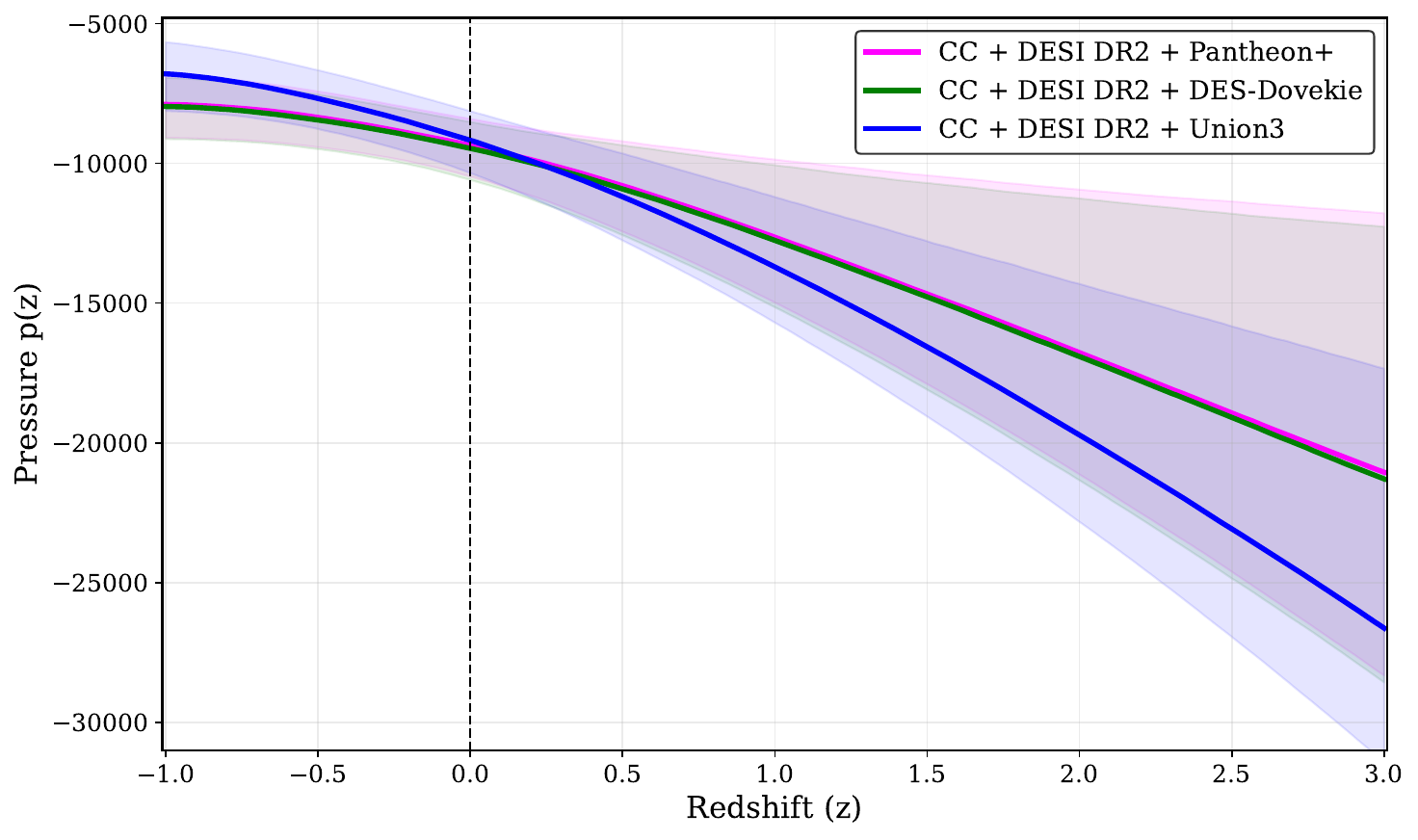}
  \caption{\justifying Evolution of the effective pressure constrained by the CC + DESI DR2 data combined separately with the Pantheon+, DES-Dovekie and Union3 Type Ia supernova datasets.}
  \label{fig:3.11}
  \end{figure}

Throughout the cosmic evolution, the effective energy density, $\rho$, remains positive and increases with increasing redshift, as shown in Fig.~\ref{fig:3.10}. It approaches zero in the far-future limit $z \to -1$, while remaining positive throughout the considered redshift range. In contrast, the effective pressure, $p$, shown in Fig.~\ref{fig:3.11}, remains negative and decreases in magnitude towards the present epoch. The emergence of a negative effective pressure is a characteristic feature associated with dark energy driven cosmic acceleration. Therefore, the evolution of the effective pressure predicted by the model exhibits the behavior required to drive the observed late-time cosmic acceleration. These behaviors are obtained for the parameter values constrained by the three combined observational datasets, providing further support for the consistency of the model with observational data.

Moreover, the total (or effective) equation of state (EoS) parameter, defined as \(\omega=p/\rho\), provides a useful characterization of the cosmic expansion history. Conventionally, \(\omega=1\), \(\omega=1/3\) and \(\omega=0\) correspond to stiff-fluid, radiation-dominated and pressureless-matter-dominated regimes, respectively. Combining Eqs.~\eqref{eq:3.22} and \eqref{eq:3.45}, the effective EoS parameter and the deceleration parameter are related by
\begin{equation}\label{eq:3.50}
\omega(z)=-1+\frac{2(m+1)}{3}[1+q(z)].
\end{equation}
Equivalently,
\begin{equation}\label{eq:3.51}
q(z)=\frac{3[1+\omega(z)]}{2(m+1)}-1.
\end{equation}
For $m>-1$, the condition for accelerated expansion, \(q(z)<0\), becomes
\begin{equation*}
\omega<\frac{2m-1}{3}.
\end{equation*}
Thus, the condition for accelerated expansion depends on the model parameter \(m\) and recovers the standard GR condition \(\omega<-1/3\) for $m=0$.

\begin{figure}
\renewcommand{\figurename}{Fig.}
\centering
\includegraphics[width=0.74\textwidth]{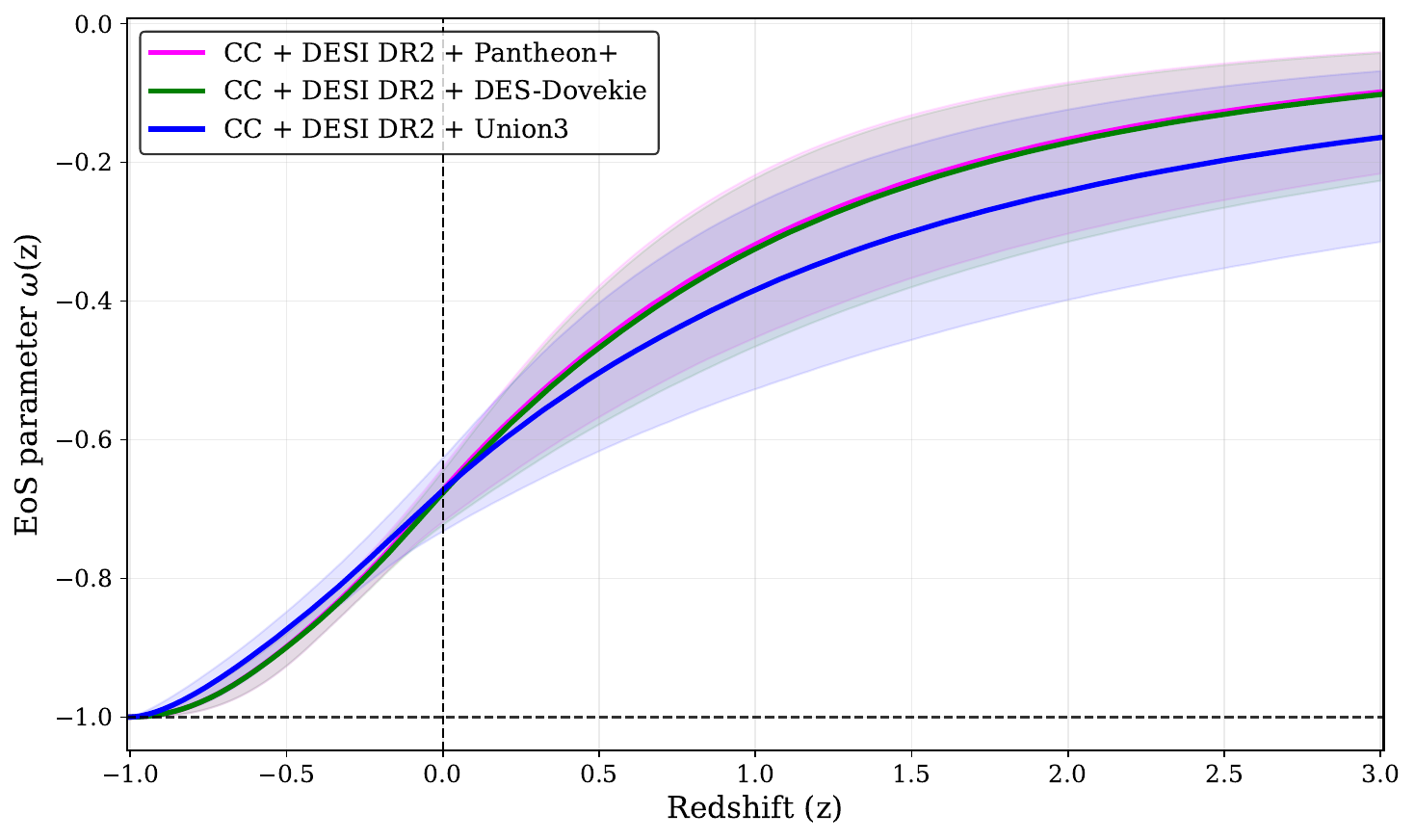}
 \caption{\justifying Evolution of the effective equation of state parameter constrained by the CC + DESI DR2 data combined separately with the Pantheon+, DES-Dovekie and Union3 Type Ia supernova datasets.}
  \label{fig:3.12}
  \end{figure}

Fig.~\ref{fig:3.12} shows the evolution of the effective EoS parameter for the CC + DESI DR2 + Pantheon+, CC + DESI DR2 + DES-Dovekie and CC + DESI DR2 + Union3 dataset combinations. The corresponding present-day values are \(\omega_0=-0.673^{+0.033}_{-0.046}\), \(\omega_0=-0.677^{+0.031}_{-0.047}\) and \(\omega_0=-0.673^{+0.046}_{-0.059}\), respectively. 
These values lie in the interval \(-1<\omega_0<-1/3\), indicating a negative effective pressure relative to the total energy density at the present epoch. In the present model, accelerated expansion requires \(\omega_0<(2m-1)/3\) for \(m>-1\). Using the best-fit values of \(m\), we find that this condition is satisfied for all three dataset combinations, consistent with accelerated expansion at the present epoch.

\section{Observationally constrained cosmic dynamics in $f(Q)$ gravity: Statefinder and $Om(z)$ diagnostics}\label{sec:3.07}
To gain deeper insight into the nature of cosmic acceleration and to distinguish between different cosmological scenarios, particularly those associated with dark energy, the Statefinder diagnostic was introduced as a valuable geometrical tool. It is constructed from the higher-order time derivatives of the cosmic scale factor and therefore provides a geometrical characterization of the expansion dynamics of the universe. Unlike diagnostics based directly on the energy density and pressure of dark energy, which may become model-dependent in modified theories of gravity, the Statefinder diagnostic is formulated purely in terms of the cosmic expansion history. The dimensionless cosmological diagnostic pair $\{r,s\}$, introduced by~\citet{Sahni_2003} and further explored by~\citet{Alam_2003}, is constructed from the scale factor $a(t)$ and its time derivatives and is therefore directly related to the underlying space-time geometry. This makes the Statefinder diagnostic particularly useful for differentiating between various dark energy models and comparing their cosmological evolution.

The Statefinder parameters are defined as
\begin{equation}\label{eq:3.52}
r=\frac{\dddot{a}}{aH^3},
\qquad
s=\frac{r-1}{3\left(q-\frac{1}{2}\right)}.
\end{equation}
The $\Lambda$CDM model corresponds to the fixed point $(r,s)=(1,0)$ in the Statefinder plane. In general, the region characterized by $r>1$ and $s<0$ is associated with Chaplygin gas-like behavior, whereas $r<1$ and $s>0$ corresponds to quintessence-like behavior.

\begin{figure}
\renewcommand{\figurename}{Fig.}
\centering
\includegraphics[width=0.74\textwidth]{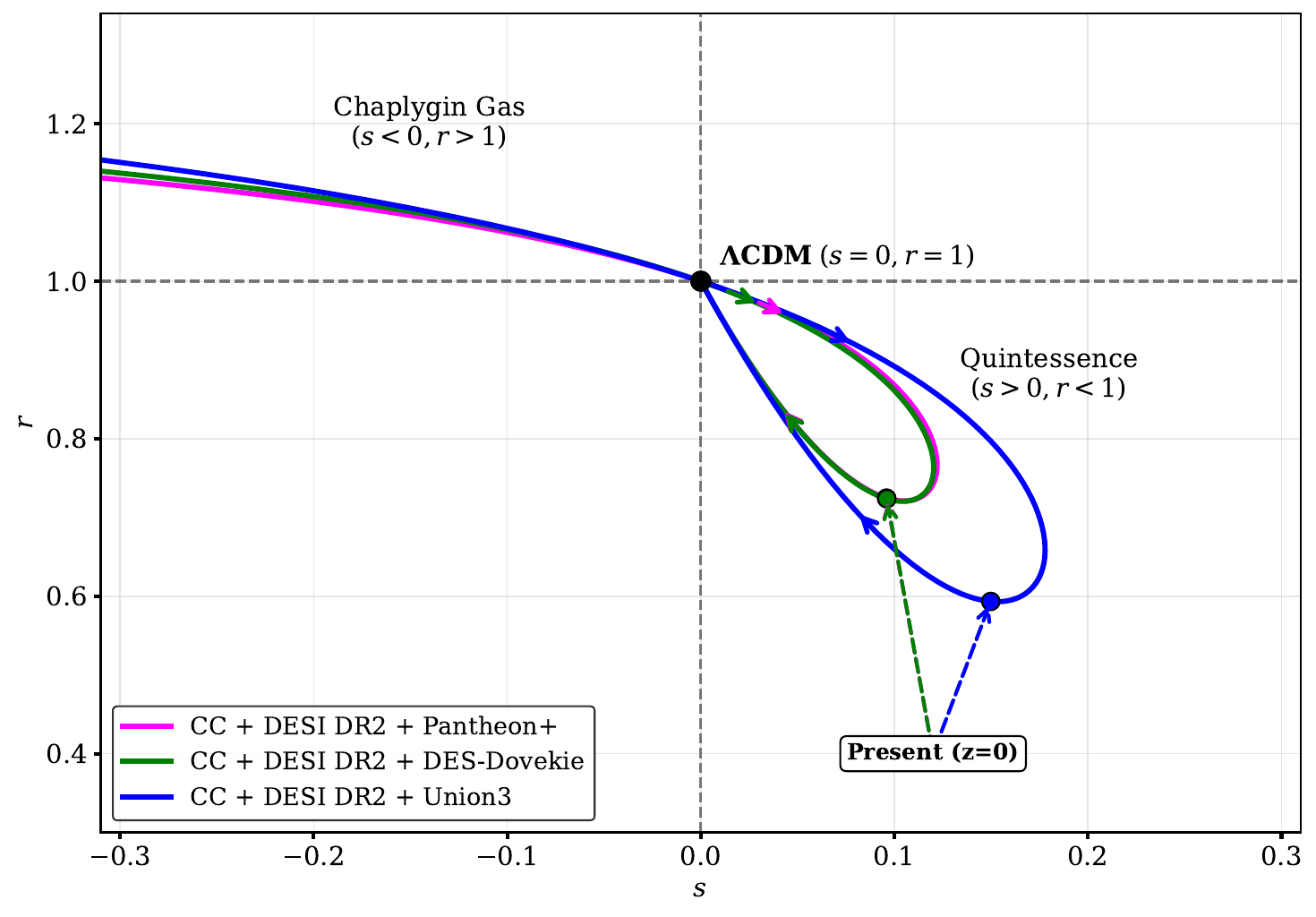}
  \caption{\justifying Statefinder diagnostics as a function of redshift $z$ for the $f(Q)$ model constrained by the CC + DESI DR2 data combined separately with the Pantheon+, DES-Dovekie and Union3 Type Ia supernova datasets.}
  \label{fig:3.13}
  \end{figure}

The evolutionary trajectories of the Statefinder parameters $(s,r)$ under observational constraints are presented in Fig.~\ref{fig:3.13}. The trajectories evolve within the quintessence-like region, characterized by $r<1$ and $s>0$ and subsequently approach the $\Lambda$CDM fixed point, corresponding to $(s,r)=(0,1)$,
in the far future limit. This behavior indicates that the model exhibits quintessence-like characteristics during its evolution while asymptotically approaching $\Lambda$CDM behavior in the distant future.

The $Om(z)$ diagnostic provides a useful, largely model-independent tool for characterizing the late-time expansion history and distinguishing between different dark energy scenarios~\citep{Sahni_2008}. A key advantage of this diagnostic is its simplicity, as it depends only on the Hubble parameter and does not require higher derivatives of the scale factor. For a spatially flat universe, it is defined as
\begin{equation}\label{eq:3.53}
Om(z)=\frac{\left[\dfrac{H(z)}{H_0}\right]^2-1}{(1+z)^3-1}.
\end{equation}
For the spatially flat $\Lambda$CDM model, $Om(z)$ remains constant and is equal to the present-day matter density parameter $\Omega_{m0}$, corresponding to an effective equation of state parameter $\omega=-1$. A decreasing $Om(z)$ with redshift is generally associated with a quintessence-like regime, characterized by $\omega>-1$, whereas an increasing trend indicates phantom-like behavior with $\omega<-1$.

\begin{figure}
\renewcommand{\figurename}{Fig.}
\centering
\includegraphics[width=0.74\textwidth]{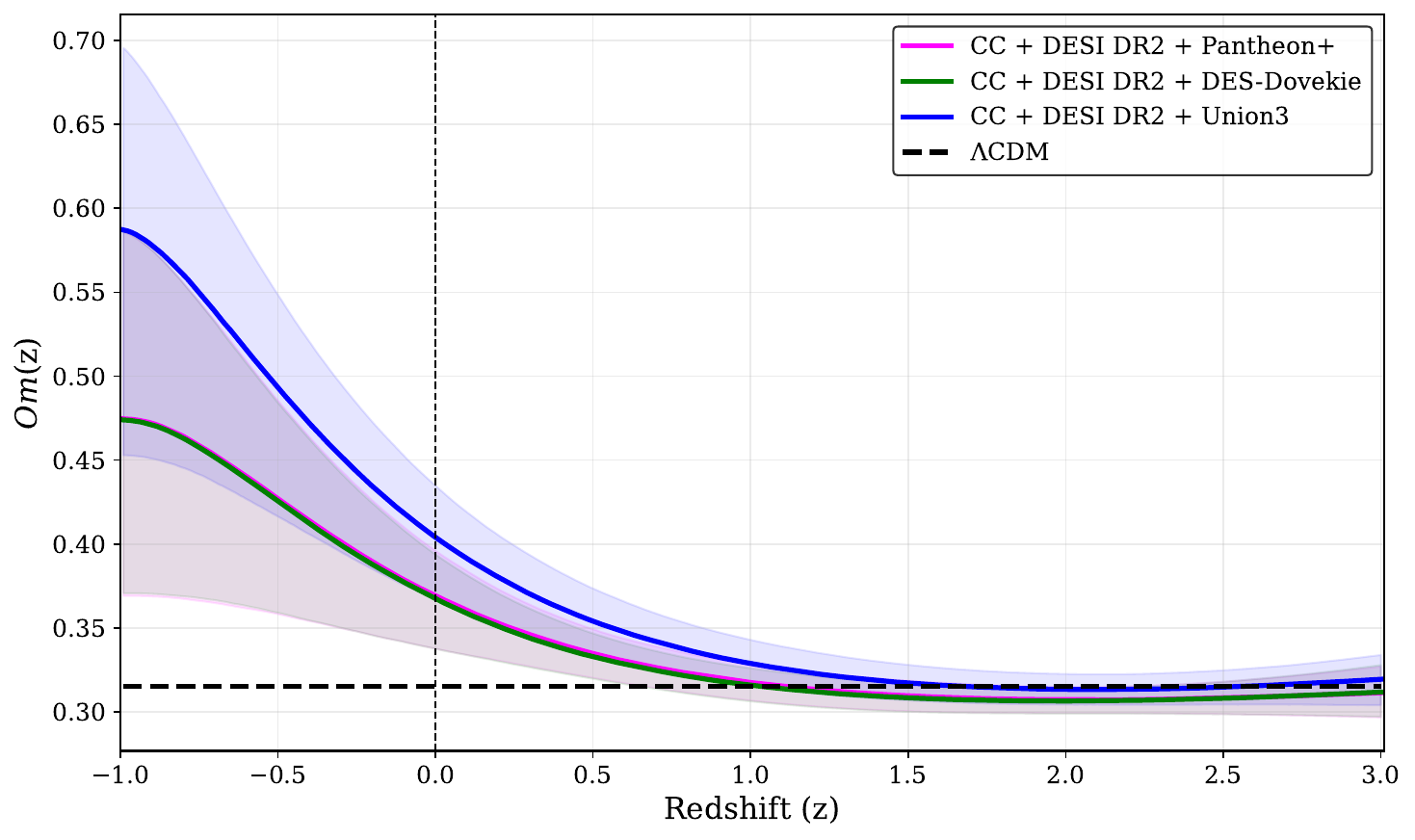}
  \caption{\justifying $Om(z)$ diagnostic as a function of redshift $z$ for the $f(Q)$ model compared to the standard $\Lambda\text{CDM}$ cosmology, constrained by the CC + DESI DR2 data combined separately with the Pantheon+, DES-Dovekie and Union3 Type Ia supernova datasets.}
  \label{fig:3.14}
  \end{figure}

As presented in Fig.~\ref{fig:3.14}, the $Om(z)$ diagnostic decreases continuously over the considered redshift range for all observational data combinations. This behavior indicates a quintessence-like regime, in agreement with the results obtained from the Statefinder diagnostics. The overall downward trend of $Om(z)$ with redshift is also qualitatively consistent with the late-time evolution from a matter-dominated, decelerating phase toward an accelerated epoch dominated by the effective dark energy component. The consistency of this behavior across the different observational data combinations further demonstrates the utility of the $Om(z)$ diagnostic in assessing the consistency of the reconstructed cosmic expansion history.

\section{Concluding remarks}\label{sec:3.08}
The growing precision and complementarity of late-time cosmological observations provide an increasingly stringent test of modified-gravity scenarios and their ability to reproduce the observed expansion history of the universe across independent datasets. In this work, we have investigated the late-time cosmic evolution within the framework of symmetric teleparallel $f(Q)$ gravity by adopting a power-law form, $f(Q)=\beta Q^{m+1}$ and characterizing the resulting dynamics through a phenomenological effective equation of state parameter parametrization. 
By combining CC measurements and DESI DR2 BAO observations with three independent Type Ia supernova compilations, namely Pantheon+, DES-Dovekie and Union3, we have constrained the model parameters and examined the reconstructed expansion history, kinematical quantities and physical properties of the model.

The joint analysis yields consistent present-day expansion rates of $H_0=68.0^{+2.7}_{-3.3}$, $68.2^{+2.7}_{-3.3}$ and $67.4^{+2.7}_{-3.3}{}$ $\mathrm{km\,s^{-1}\,Mpc^{-1}}$ for the CC + DESI DR2 + Pantheon+, CC + DESI DR2 + DES-Dovekie and CC + DESI DR2 + Union3 combinations, respectively. These estimates have substantially overlapping $1\sigma$ credible intervals and are consistent, within their uncertainties, with the value of $H_0$ inferred from Planck 2018 observations assuming the standard $\Lambda$CDM cosmology. The reconstructed $H(z)$ histories and BAO distance ratios remain broadly consistent with the flat $\Lambda$CDM prediction, while the inferred sound horizon at the drag epoch has central values of approximately $146.5$--$146.7\,\mathrm{Mpc}$. The strong anticorrelation between $H_0$ and $r_d$ reflects their degeneracy in BAO-based cosmological constraints.

The statistical comparison provides a mixed assessment of the proposed $f(Q)$ model relative to flat $\Lambda$CDM. The $f(Q)$ model yields lower minimum $\chi^2$ values and is favored by the AIC for all three dataset combinations, with $\Delta\mathrm{AIC}=-1.84$, $-2.18$ and $-5.29$ for CC + DESI DR2 + Pantheon+, CC + DESI DR2 + DES-Dovekie and CC + DESI DR2 + Union3, respectively. In contrast, the BIC favors $\Lambda$CDM for CC + DESI DR2 + Pantheon+ and CC + DESI DR2 + DES-Dovekie, with $\Delta\mathrm{BIC}=+8.99$ and $+8.89$, while the CC + DESI DR2 + Union3 result, $\Delta\mathrm{BIC}=-0.90$, indicates essentially no meaningful preference between the models. Thus, despite its improved best-fit values and favorable AIC results, the proposed $f(Q)$ model does not show a consistent statistical preference over $\Lambda$CDM when its additional parameter complexity is taken into account.

The deceleration parameter indicates a transition from decelerated to accelerated expansion, with present-day values of $q_0=-0.446^{+0.040}_{-0.047}$, $-0.449^{+0.040}_{-0.045}$ and $-0.394^{+0.046}_{-0.055}$ for the three dataset combinations. The corresponding transition redshifts lie within the narrow range $0.695 \leq z_t \leq 0.700$, indicating that the inferred onset of late-time acceleration is consistent across the independent supernova compilations and is relatively insensitive to the choice of SNe Ia dataset. The higher-order kinematical diagnostics provide further insight into possible departures from the standard cosmological expansion history. The present-day jerk parameter values lie below the flat $\Lambda$CDM prediction, $j_0=1$, indicating differences in the reconstructed kinematical evolution, although the statistical significance of these differences depends on the associated uncertainties. The present-day snap parameter values are also relatively consistent across the three reconstructions, whereas their evolution exhibits greater sensitivity to the choice of dataset at higher redshifts. Overall, the principal late-time quantities, particularly $q_0$ and $z_t$, are comparatively robust across the datasets, whereas higher-order cosmographic quantities are more sensitive to differences in the redshift evolution of the reconstructed expansion rate.

The effective energy density remains positive throughout the considered redshift range, while the effective pressure remains negative, with its magnitude decreasing towards the present epoch. This pressure behavior is consistent with the negative pressure associated with late-time cosmic acceleration. These features are consistently obtained across all three observational dataset combinations. The effective EoS parameter, which characterizes the total cosmic fluid, has present-day values of $\omega_0=-0.673^{+0.033}_{-0.046}$, $-0.677^{+0.031}_{-0.047}$ and $-0.673^{+0.046}_{-0.059}$ for the CC + DESI DR2 + Pantheon+, CC + DESI DR2 + DES-Dovekie and CC + DESI DR2 + Union3 dataset combinations, respectively. Furthermore, the best-fit parameters satisfy the model-dependent acceleration condition, $\omega_0<(2m-1)/3$, for all three combinations, consistent with accelerated expansion at the present epoch.

Furthermore, the Statefinder and $Om(z)$ diagnostics provide complementary insights into the reconstructed cosmic expansion history. The Statefinder trajectories evolve through the quintessence-like region and approach the $\Lambda$CDM fixed point, $(s,r)=(0,1)$, in the far future. Meanwhile, the decreasing $Om(z)$ diagnostic indicates quintessence-like behavior, complementing the results of the Statefinder analysis. These trends are consistent with the reconstructed evolution toward late-time accelerated expansion across all three observational dataset combinations.

Overall, the adopted EoS parametrization within the power-law $f(Q)$ framework provides an observationally viable description of late-time cosmic acceleration for the datasets considered. The reconstructed deceleration parameter indicates a transition from decelerated to accelerated expansion, while the effective EoS characterizes the evolution of the total cosmic fluid, with its negative present-day pressure consistent with the observed late-time acceleration. The consistency of the reconstructed expansion history and transition redshift across the three dataset combinations indicates that these results are robust to the choice of SNe Ia compilation when combined with the same CC and DESI DR2 BAO data. However, the model-selection analysis does not establish a consistent statistical preference for the proposed $f(Q)$ model over flat $\Lambda$CDM. Future, more precise BAO, supernova and expansion-rate measurements will help determine whether the additional parameter of the proposed $f(Q)$ model is justified by cosmological observations.

\section*{Declaration of competing interest}
The authors declare that they have no known competing financial
interests or personal relationships that could have appeared to
influence the work reported in this paper.

\section*{Data availability}
No data was used for the research described in the article.

\bibliographystyle{apsrev4-2}

\bibliography{References3}

@article{Riess_1998,
doi = {https://doi.org/10.1086/300499},
year = {1998},
publisher = {},
volume = {116},
number = {3},
pages = {1009},
author = {Riess, Adam G. and Filippenko, Alexei V. and Challis, Peter and Clocchiatti, Alejandro and Diercks, Alan and Garnavich, Peter M. and Gilliland, Ron L. and Hogan, Craig J. and Jha, Saurabh and Kirshner, Robert P. and others},
title = {{Observational Evidence from Supernovae for an Accelerating Universe and a Cosmological Constant}},
journal = {The Astronomical Journal}
}

@article{Perlmutter_1999,
doi = {https://doi.org/10.1086/307221},
year = {1999},
publisher = {},
volume = {517},
number = {2},
pages = {565},
author = {Perlmutter, S. and Aldering, G. and Goldhaber, G. and Knop, R. A. and Nugent, P. and Castro, P. G. and Deustua, S. and Fabbro, S. and Goobar, A. and Groom, D. E. and others},
title = {{Measurements of $\Omega$ and $\Lambda$ from 42 High-Redshift Supernovae}},
journal = {The Astrophysical Journal}
}

@article{Riess_2004,
doi = {https://doi.org/10.1086/383612},
year = {2004},
publisher = {},
volume = {607},
number = {2},
pages = {665},
author = {Riess, Adam G. and Strolger, Louis-Gregory and Tonry, John and Casertano, Stefano and Ferguson, Henry C. and Mobasher, Bahram and Challis, Peter and Filippenko, Alexei V. and Jha, Saurabh and Li, Weidong and others},
title = {{Type Ia Supernova Discoveries at $z>1$ from the Hubble Space Telescope: Evidence for Past Deceleration and Constraints on Dark Energy Evolution}},
journal = {The Astrophysical Journal}
}

@article{Caldwell_2004,
  title = {{Cosmic microwave background and supernova constraints on quintessence: Concordance regions and target models}},
  author = {Caldwell, Robert R. and Doran, Michael},
  journal = {Phys. Rev. D},
  volume = {69},
  issue = {10},
  pages = {103517},
  numpages = {6},
  year = {2004},
  publisher = {American Physical Society},
  doi = {https://doi.org/10.1103/PhysRevD.69.103517}
}

@article{Eisenstein_2005,
doi = {https://doi.org/10.1086/466512},
year = {2005},
publisher = {},
volume = {633},
number = {2},
pages = {560},
author = {Eisenstein, Daniel J. and Zehavi, Idit and Hogg, David W. and Scoccimarro, Roman and Blanton, Michael R. and Nichol, Robert C. and Scranton, Ryan and Seo, Hee-Jong and Tegmark, Max and Zheng, Zheng and others},
title = {{Detection of the Baryon Acoustic Peak in the Large-Scale Correlation Function of SDSS Luminous Red Galaxies}},
journal = {The Astrophysical Journal}
}

@article{Percival_2010,
    author = {Percival, Will J. and Reid, Beth A. and Eisenstein, Daniel J. and Bahcall, Neta A. and Budavari, Tamas and Frieman, Joshua A. and Fukugita, Masataka and Gunn, James E. and Ivezi\'{c}, {\v{Z}}eljko
    and Knapp, Gillian R. and others},
    title = {{Baryon acoustic oscillations in the Sloan Digital Sky Survey Data Release 7 galaxy sample}},
    journal = {Monthly Notices of the Royal Astronomical Society},
    volume = {401},
    number = {4},
    pages = {2148-2168},
    year = {2010},
    doi = {https://doi.org/10.1111/j.1365-2966.2009.15812.x}
}

@article{Koivisto_2006,
  title = {{Dark energy anisotropic stress and large scale structure formation}},
  author = {Koivisto, Tomi and Mota, David F.},
  journal = {Phys. Rev. D},
  volume = {73},
  issue = {8},
  pages = {083502},
  numpages = {12},
  year = {2006},
  publisher = {American Physical Society},
  doi = {https://doi.org/10.1103/PhysRevD.73.083502}
}

@article{Daniel_2008,
  title = {{Large scale structure as a probe of gravitational slip}},
  author = {Daniel, Scott F. and Caldwell, Robert R. and Cooray, Asantha and Melchiorri, Alessandro},
  journal = {Phys. Rev. D},
  volume = {77},
  issue = {10},
  pages = {103513},
  numpages = {12},
  year = {2008},
  publisher = {American Physical Society},
  doi = {https://doi.org/10.1103/PhysRevD.77.103513}
}

@article{Spergel_2003,
doi = {https://doi.org/10.1086/377226},
year = {2003},
publisher = {},
volume = {148},
number = {1},
pages = {175},
author = {Spergel, D. N. and Verde, L. and Peiris, H. V. and Komatsu, E. and Nolta, M. R. and Bennett, C. L. and Halpern, M. and Hinshaw, G. and Jarosik, N. and Kogut, A. and others},
title = {{First-Year Wilkinson Microwave Anisotropy Probe
(WMAP) Observations:
Determination of Cosmological Parameters}},
journal = {The Astrophysical Journal Supplement Series}
}

@article{Abdul_2025,
  title = {{DESI DR2 results. II. Measurements of baryon acoustic oscillations and cosmological constraints}},
  author = {Abdul Karim, M. and Aguilar, J. and Ahlen, S. and Alam, S. and Allen, L. and Prieto, C. Allende and Alves, O. and Anand, A. and Andrade, U. and Armengaud, E. and Aviles, A. and others},
  journal = {Phys. Rev. D},
  volume = {112},
  issue = {8},
  pages = {083515},
  numpages = {40},
  year = {2025},
  publisher = {American Physical Society},
  doi = {https://doi.org/10.1103/tr6y-kpc6}
}

@article{Lodha_2025,
  title = {{Extended dark energy analysis using DESI DR2 BAO measurements}},
  author = {Lodha, K. and Calderon, R. and Matthewson, W. L. and Shafieloo, A. and Ishak, M. and Pan, J. and Garcia-Quintero, C. and Huterer, D. and Valogiannis, G. and Ure\~na-L\'opez, L. A. and others},
  journal = {Phys. Rev. D},
  volume = {112},
  issue = {8},
  pages = {083511},
  numpages = {27},
  year = {2025},
  publisher = {American Physical Society},
  doi = {https://doi.org/10.1103/w4c6-1r5j}
}

@article{Einstein_1916,
author = {Einstein, A.},
title = {{Die Grundlage der allgemeinen Relativitätstheorie}},
journal = {Annalen der Physik},
volume = {354},
number = {7},
pages = {769-822},
doi = {https://doi.org/10.1002/andp.19163540702},
year = {1916}
}

@article{Weinberg_1989,
  title = {{The cosmological constant problem}},
  author = {Weinberg, Steven},
  journal = {Rev. Mod. Phys.},
  volume = {61},
  issue = {1},
  pages = {1--23},
  numpages = {0},
  year = {1989},
  publisher = {American Physical Society},
  doi = {https://doi.org/10.1103/RevModPhys.61.1}
}

@article{Dalal_2001,
  title = {{Testing the Cosmic Coincidence Problem and the Nature of Dark Energy}},
  author = {Dalal, Neal and Abazajian, Kevork and Jenkins, Elizabeth and Manohar, Aneesh V.},
  journal = {Phys. Rev. Lett.},
  volume = {87},
  issue = {14},
  pages = {141302},
  numpages = {4},
  year = {2001},
  publisher = {American Physical Society},
  doi = {https://doi.org/10.1103/PhysRevLett.87.141302}
}

@article{Verde_2019,
  title={{Tensions between the early and late Universe}},
  author={Verde, Licia and Treu, Tommaso and Riess, Adam G},
  journal={Nature Astronomy},
  volume={3},
  number={10},
  pages={891--895},
  year={2019},
  doi={https://doi.org/10.1038/s41550-019-0902-0},
  publisher={Nature Publishing Group UK London}
}

@article{Valentino_2021,
doi = {https://doi.org/10.1088/1361-6382/ac086d},
year = {2021},
publisher = {IOP Publishing},
volume = {38},
number = {15},
pages = {153001},
author = {Di Valentino, Eleonora and Mena, Olga and Pan, Supriya and Visinelli, Luca and Yang, Weiqiang and Melchiorri, Alessandro and Mota, David F and Riess, Adam G and Silk, Joseph},
title = {{In the realm of the Hubble tension—a review of solutions}},
journal = {Classical and Quantum Gravity}
}

@article{Copeland_2006,
author = {COPELAND, EDMUND J. and SAMI, M. and TSUJIKAWA, SHINJI},
title = {{DYNAMICS OF DARK ENERGY}},
journal = {International Journal of Modern Physics D},
volume = {15},
number = {11},
pages = {1753-1935},
year = {2006},
doi = {https://doi.org/10.1142/S021827180600942X}
}

@article{Armendariz_2001,
  title = {{Essentials of k-essence}},
  author = {Armendariz-Picon, C. and Mukhanov, V. and Steinhardt, Paul J.},
  journal = {Phys. Rev. D},
  volume = {63},
  issue = {10},
  pages = {103510},
  numpages = {13},
  year = {2001},
  publisher = {American Physical Society},
  doi = {https://doi.org/10.1103/PhysRevD.63.103510}
}

@article{Bagla_2003,
  title = {{Cosmology with tachyon field as dark energy}},
  author = {Bagla, J. S. and Jassal, H. K. and Padmanabhan, T.},
  journal = {Phys. Rev. D},
  volume = {67},
  issue = {6},
  pages = {063504},
  numpages = {11},
  year = {2003},
  publisher = {American Physical Society},
  doi = {https://doi.org/10.1103/PhysRevD.67.063504}
}

@article{Bento_2002,
  title = {{Generalized Chaplygin gas, accelerated expansion, and dark-energy-matter unification}},
  author = {Bento, M. C. and Bertolami, O. and Sen, A. A.},
  journal = {Phys. Rev. D},
  volume = {66},
  issue = {4},
  pages = {043507},
  numpages = {5},
  year = {2002},
  publisher = {American Physical Society},
  doi = {https://doi.org/10.1103/PhysRevD.66.043507}
}

@article{Debnath_2004,
doi = {https://doi.org/10.1088/0264-9381/21/23/019},
year = {2004},
publisher = {},
volume = {21},
number = {23},
pages = {5609},
author = {Debnath, Ujjal and Banerjee, Asit and Chakraborty, Subenoy},
title = {{Role of modified Chaplygin gas in accelerated universe}},
journal = {Classical and Quantum Gravity}
}

@article{Buchdahl_1970,
    author = {Buchdahl, H. A.},
    title = {{Non-Linear Lagrangians and Cosmological Theory}},
    journal = {Monthly Notices of the Royal Astronomical Society},
    volume = {150},
    number = {1},
    pages = {1-8},
    year = {1970},
    doi = {https://doi.org/10.1093/mnras/150.1.1}
}

@article{Starobinsky_1980,
title = {{A new type of isotropic cosmological models without singularity}},
journal = {Physics Letters B},
volume = {91},
number = {1},
pages = {99-102},
year = {1980},
doi = {https://doi.org/10.1016/0370-2693(80)90670-X},
author = {A.A. Starobinsky}
}

@article{Harko_2011,
  title = {{$f(R,T)$ gravity}},
  author = {Harko, Tiberiu and Lobo, Francisco S. N. and Nojiri, Shin'ichi and Odintsov, Sergei D.},
  journal = {Phys. Rev. D},
  volume = {84},
  issue = {2},
  pages = {024020},
  numpages = {11},
  year = {2011},
  publisher = {American Physical Society},
  doi = {https://doi.org/10.1103/PhysRevD.84.024020}
}

@article{Harko_2010,
  title={{$f(R,L_m)$ gravity}},
  author={Harko, Tiberiu and Lobo, Francisco SN},
  journal={The European Physical Journal C},
  volume={70},
  number={1-2},
  pages={373--379},
  year={2010},
doi={https://doi.org/10.1140/epjc/s10052-010-1467-3},
  publisher={Springer}
}

@article{Ferraro_2007,
  title = {{Modified teleparallel gravity: Inflation without an inflaton}},
  author = {Ferraro, Rafael and Fiorini, Franco},
  journal = {Phys. Rev. D},
  volume = {75},
  issue = {8},
  pages = {084031},
  numpages = {5},
  year = {2007},
  publisher = {American Physical Society},
  doi = {https://doi.org/10.1103/PhysRevD.75.084031}
}

@article{Jimenez_2018,
  title = {{Coincident general relativity}},
  author = {Jim\'enez, Jose Beltr\'an and Heisenberg, Lavinia and Koivisto, Tomi},
  journal = {Phys. Rev. D},
  volume = {98},
  issue = {4},
  pages = {044048},
  numpages = {6},
  year = {2018},
  publisher = {American Physical Society},
  doi = {https://doi.org/10.1103/PhysRevD.98.044048}
}

@article{Heisenberg_2024,
title = {{Review on $f(Q)$ gravity}},
journal = {Physics Reports},
volume = {1066},
pages = {1-78},
year = {2024},
doi = {https://doi.org/10.1016/j.physrep.2024.02.001},
author = {Lavinia Heisenberg}
}

@article{Shabani_2024,
  title={{Cosmology of $f(Q)$ gravity in non-flat Universe}},
  author={Shabani, Hamid and De, Avik and Loo, Tee-How and Saridakis, Emmanuel N},
  journal={The European Physical Journal C},
  volume={84},
  number={3},
  pages={285},
  year={2024},
  doi={https://doi.org/10.1140/epjc/s10052-024-12582-3},
  publisher={Springer}
}

@article{Mandal_2020,
  title = {{Energy conditions in $f(Q)$ gravity}},
  author = {Mandal, Sanjay and Sahoo, P. K. and Santos, J. R. L.},
  journal = {Phys. Rev. D},
  volume = {102},
  issue = {2},
  pages = {024057},
  numpages = {8},
  year = {2020},
  publisher = {American Physical Society},
  doi = {https://doi.org/10.1103/PhysRevD.102.024057}
}

@article{Solanki_2021,
title = {{Cosmic acceleration with bulk viscosity in modified $f(Q)$ gravity}},
journal = {Physics of the Dark Universe},
volume = {32},
pages = {100820},
year = {2021},
doi = {https://doi.org/10.1016/j.dark.2021.100820},
author = {Raja Solanki and S K J. Pacif and Abhishek Parida and P K. Sahoo}
}

@article{Singh_2026Observational,
title = {{Observational constraints on cosmic evolution in $f(Q)$ gravity using logarithmic $Om(z)$ parametrization}},
journal = {The European Physical Journal Plus},
volume = {141},
pages = {971},
year = {2026},
doi = {https://doi.org/10.1140/epjp/s13360-026-08200-8},
author = {Singh, Kshetrimayum Govind and Singh, Kangujam Priyokumar}
}

@article{Arora_2026,
title = {{Bayesian and machine-learning analyses of nonminimal $f(Q)$ gravity and $H_0$ tension}},
journal = {Journal of High Energy Astrophysics},
volume = {54},
pages = {100682},
year = {2026},
doi = {https://doi.org/10.1016/j.jheap.2026.100682},
author = {Simran Arora and Mridul Patel}
}

@article{Nashed_2026,
title = {{Late-time cosmology and structure formation in quadratic $f(Q)$ gravity}},
journal = {Physics Letters B},
volume = {878},
pages = {140575},
year = {2026},
doi = {https://doi.org/10.1016/j.physletb.2026.140575},
author = {G.G.L. Nashed and P.V. Tretyakov and A. Eid}
}

@article{Paliathanasis_2026,
title = {{Observational constraints on noncoincident $f(Q)$-gravity with matter-gravity coupling}},
journal = {Journal of High Energy Astrophysics},
volume = {53},
pages = {100609},
year = {2026},
doi = {https://doi.org/10.1016/j.jheap.2026.100609},
author = {Andronikos Paliathanasis}
}

@article{Kolhatkar_2026,
title = {{Beyond the cosmological constant: Breaking the geometric degeneracy of $f(Q)$ cosmology via redshift-space distortions}},
journal = {Physics Letters B},
volume = {879},
pages = {140709},
year = {2026},
doi = {https://doi.org/10.1016/j.physletb.2026.140709},
author = {Ameya Kolhatkar and P K. Sahoo}
}

@article{Mazumdar_2026,
doi = {https://doi.org/10.1088/1361-6382/ae5cf4},
year = {2026},
publisher = {IOP Publishing},
volume = {43},
number = {8},
pages = {085004},
author = {Mazumdar, Rajdeep and Malakar, Kalyan and Bhuyan, Kalyan},
title = {{Fractional holographic dark energy driven reconstruction of $f(Q)$ gravity and its cosmological implications}},
journal = {Classical and Quantum Gravity}
}

@article{Chakraborty_2025,
doi = {https://doi.org/10.1088/1475-7516/2025/05/098},
year = {2025},
publisher = {IOP Publishing},
volume = {2025},
number = {05},
pages = {098},
author = {Chakraborty, Saikat and Dutta, Jibitesh and Gregoris, Daniele and Karwan, Khamphee and Khyllep, Wompherdeiki},
title = {{Reproducing $\Lambda$CDM-like solutions in $f(Q)$ gravity: a comprehensive study across all connection branches}},
journal = {Journal of Cosmology and Astroparticle Physics}
}

@article{Lymperis_2022,
doi = {https://doi.org/10.1088/1475-7516/2022/11/018},
year = {2022},
publisher = {IOP Publishing},
volume = {2022},
number = {11},
pages = {018},
author = {Lymperis, Andreas},
title = {{Late-time cosmology with phantom dark-energy in $f(Q)$ gravity}},
journal = {Journal of Cosmology and Astroparticle Physics}
}

@article{Narawade_2023,
title = {{Accelerating cosmological models in $f(Q)$ gravity and the phase space analysis}},
journal = {Physics of the Dark Universe},
volume = {42},
pages = {101282},
year = {2023},
issn = {2212-6864},
doi = {https://doi.org/10.1016/j.dark.2023.101282},
author = {S.A. Narawade and Shashank P. Singh and B. Mishra}
}

@article{Koussour_2022,
title = {{Cosmic acceleration and energy conditions in symmetric teleparallel $f(Q)$ gravity}},
journal = {Journal of High Energy Astrophysics},
volume = {35},
pages = {43-51},
year = {2022},
issn = {2214-4048},
doi = {https://doi.org/10.1016/j.jheap.2022.05.002},
author = {M. Koussour and S.H. Shekh and M. Bennai}
}

@article{Dubey_2025,
title = {{Study of cosmological dark energy models under $f(Q)$ gravity}},
journal = {Physics of the Dark Universe},
volume = {47},
pages = {101736},
year = {2025},
issn = {2212-6864},
doi = {https://doi.org/10.1016/j.dark.2024.101736},
author = {Vipin Chandra Dubey and Umesh Kumar Sharma and Saibal Ray and Aritra Sanyal}
}

@article{Yadav_2024,
title = {{Reconstructing $f(Q)$ gravity from parameterization of the Hubble parameter and observational constraints}},
journal = {Journal of High Energy Astrophysics},
volume = {43},
pages = {114-125},
year = {2024},
issn = {2214-4048},
doi = {https://doi.org/10.1016/j.jheap.2024.06.012},
author = {Anil Kumar Yadav and S.R. Bhoyar and M.C. Dhabe and S.H. Shekh and Nafis Ahmad}
}

@article{Lazkoz_2019,
  title = {{Observational constraints of $f(Q)$ gravity}},
  author = {Lazkoz, Ruth and Lobo, Francisco S. N. and Ortiz-Ba\~nos, Mar\'{\i}a and Salzano, Vincenzo},
  journal = {Phys. Rev. D},
  volume = {100},
  issue = {10},
  pages = {104027},
  numpages = {8},
  year = {2019},
  publisher = {American Physical Society},
  doi = {https://doi.org/10.1103/PhysRevD.100.104027}
}

@article{Mukherjee_2016,
    author = {Mukherjee, Ankan},
    title = {{Acceleration of the universe: a reconstruction of the effective equation of state}},
    journal = {Monthly Notices of the Royal Astronomical Society},
    volume = {460},
    number = {1},
    pages = {273-282},
    year = {2016},
    doi = {https://doi.org/10.1093/mnras/stw964}
}

@article{Brout_2022,
doi = {https://doi.org/10.3847/1538-4357/ac8e04},
year = {2022},
publisher = {The American Astronomical Society},
volume = {938},
number = {2},
pages = {110},
author = {Brout, Dillon and Scolnic, Dan and Popovic, Brodie and Riess, Adam G. and Carr, Anthony and Zuntz, Joe and Kessler, Rick and Davis, Tamara M. and Hinton, Samuel and Jones, David and Kenworthy, W. D’Arcy and others},
title = {{The Pantheon+ Analysis: Cosmological Constraints}},
journal = {The Astrophysical Journal}
}

@article{Scolnic_2022,
doi = {https://doi.org/10.3847/1538-4357/ac8b7a},
year = {2022},
publisher = {The American Astronomical Society},
volume = {938},
number = {2},
pages = {113},
author = {Scolnic, Dan and Brout, Dillon and Carr, Anthony and Riess, Adam G. and Davis, Tamara M. and Dwomoh, Arianna and Jones, David O. and Ali, Noor and Charvu, Pranav and Chen, Rebecca and Peterson, Erik R. and others},
title = {{The Pantheon+ Analysis: The Full Data Set and Light-curve Release}},
journal = {The Astrophysical Journal}
}

@article{Abbott_2024,
doi = {https://doi.org/10.3847/2041-8213/ad6f9f},
year = {2024},
publisher = {The American Astronomical Society},
volume = {973},
number = {1},
pages = {L14},
author = {Abbott, DES Collaboration: T. M. C. and Acevedo, M. and Aguena, M. and Alarcon, A. and Allam, S. and Alves, O. and Amon, A. and Andrade-Oliveira, F. and Annis, J. and Armstrong, P. and Asorey, J. and Avila, S. and others},
title = {{The Dark Energy Survey: Cosmology Results with $\sim$ 1500 New High-redshift Type Ia Supernovae Using the Full 5 yr Data Set}},
journal = {The Astrophysical Journal Letters}
}

@article{Popovic_2026,
    author = {Popovic, B and Shah, P and Kenworthy, W D and Kessler, R and Davis, T M and Goobar, A and Scolnic, D and Vincenzi, M and Wiseman, P and Chen, R and others},
    title = {{The Dark Energy Survey supernova program: a reanalysis of cosmology results and evidence for evolving dark energy with an updated Type Ia supernova calibration}},
    journal = {Monthly Notices of the Royal Astronomical Society},
    volume = {548},
    number = {4},
    pages = {stag632},
    year = {2026},
    doi = {https://doi.org/10.1093/mnras/stag632}
}

@article{Rubin_2025,
doi = {https://doi.org/10.3847/1538-4357/adc0a5},
year = {2025},
publisher = {The American Astronomical Society},
volume = {986},
number = {2},
pages = {231},
author = {Rubin, David and Aldering, Greg and Betoule, Marc and Fruchter, Andy and Huang, Xiaosheng and Kim, Alex G. and Lidman, Chris and Linder, Eric and Perlmutter, Saul and Ruiz-Lapuente, Pilar and Suzuki, Nao},
title = {{Union through UNITY: Cosmology with 2000 SNe Using a Unified Bayesian Framework}},
journal = {The Astrophysical Journal}
}

@article{Perivolaropoulos_2023,
    author = {Perivolaropoulos, Leandros and Skara, Foteini},
    title = {{On the homogeneity of SnIa absolute magnitude in the Pantheon+ sample}},
    journal = {Monthly Notices of the Royal Astronomical Society},
    volume = {520},
    number = {4},
    pages = {5110-5125},
    year = {2023},
    doi = {https://doi.org/10.1093/mnras/stad451}
}

@article{Foreman_2013,
doi = {https://doi.org/10.1086/670067},
year = {2013},
publisher = {University of Chicago Press},
volume = {125},
number = {925},
pages = {306},
author = {Foreman-Mackey, Daniel and Hogg, David W. and Lang, Dustin and Goodman, Jonathan},
title = {{emcee: The MCMC Hammer}},
journal = {Publications of the Astronomical Society of the Pacific}
}

@article{Ryden_2003,
  title={{Introduction to Cosmology Addison Wesley}},
  author={Ryden, B},
  journal={San Francisco, USA},
  year={2003}
}

@article{CHEVALLIER_2001,
author = {CHEVALLIER, MICHEL and POLARSKI, DAVID},
title = {{ACCELERATING UNIVERSES WITH SCALING DARK MATTER}},
journal = {International Journal of Modern Physics D},
volume = {10},
number = {02},
pages = {213-223},
year = {2001},
doi = {https://doi.org/10.1142/S0218271801000822} 
}

@article{Linder_2003,
  title = {{Exploring the Expansion History of the Universe}},
  author = {Linder, Eric V.},
  journal = {Phys. Rev. Lett.},
  volume = {90},
  issue = {9},
  pages = {091301},
  numpages = {4},
  year = {2003},
  month = {Mar},
  publisher = {American Physical Society},
  doi = {https://doi.org/10.1103/PhysRevLett.90.09130}
}

@article{Weller_2002,
  title = {{Future supernovae observations as a probe of dark energy}},
  author = {Weller, Jochen and Albrecht, Andreas},
  journal = {Phys. Rev. D},
  volume = {65},
  issue = {10},
  pages = {103512},
  numpages = {21},
  year = {2002},
  publisher = {American Physical Society},
  doi = {https://doi.org/10.1103/PhysRevD.65.103512}
}

@article{Efstathiou_1999,
    author = {Efstathiou, G.},
    title = {{Constraining the equation of state of the Universe from distant Type Ia supernovae and cosmic microwave background anisotropies}},
    journal = {Monthly Notices of the Royal Astronomical Society},
    volume = {310},
    number = {3},
    pages = {842-850},
    year = {1999},
    doi = {https://doi.org/10.1046/j.1365-8711.1999.02997.x}
}

@article{Jassal_2005,
  title = {{Observational constraints on low redshift evolution of dark energy: How consistent are different observations?}},
  author = {Jassal, H. K. and Bagla, J. S. and Padmanabhan, T.},
  journal = {Phys. Rev. D},
  volume = {72},
  issue = {10},
  pages = {103503},
  numpages = {21},
  year = {2005},
  publisher = {American Physical Society},
  doi = {https://doi.org/10.1103/PhysRevD.72.103503}
}

@article{Barboza_2008,
title = {{A parametric model for dark energy}},
journal = {Physics Letters B},
volume = {666},
number = {5},
pages = {415-419},
year = {2008},
doi = {https://doi.org/10.1016/j.physletb.2008.08.012},
author = {E.M. Barboza and J.S. Alcaniz}
}

@article{Harko_2018,
  title = {{Coupling matter in modified $Q$ gravity}},
  author = {Harko, Tiberiu and Koivisto, Tomi S. and Lobo, Francisco S. N. and Olmo, Gonzalo J. and Rubiera-Garcia, Diego},
  journal = {Phys. Rev. D},
  volume = {98},
  issue = {8},
  pages = {084043},
  numpages = {13},
  year = {2018},
  publisher = {American Physical Society},
  doi = {https://doi.org/10.1103/PhysRevD.98.084043}
}

@article{Jimenez_2002,
doi = {https://doi.org/10.1086/340549},
year = {2002},
publisher = {},
volume = {573},
number = {1},
pages = {37},
author = {Jimenez, Raul and Loeb, Abraham},
title = {{Constraining Cosmological Parameters Based on Relative Galaxy Ages}},
journal = {The Astrophysical Journal}
}

@article{Moresco_2020,
doi = {https://doi.org/10.3847/1538-4357/ab9eb0},
year = {2020},
publisher = {The American Astronomical Society},
volume = {898},
number = {1},
pages = {82},
author = {Moresco, Michele and Jimenez, Raul and Verde, Licia and Cimatti, Andrea and Pozzetti, Lucia},
title = {{Setting the Stage for Cosmic Chronometers. II. Impact of Stellar Population Synthesis Models Systematics and Full Covariance Matrix}},
journal = {The Astrophysical Journal}
}

@article{Zhang_2014,
doi = {https://doi.org/10.1088/1674-4527/14/10/002},
year = {2014},
publisher = {},
volume = {14},
number = {10},
pages = {1221},
author = {Zhang, Cong and Zhang, Han and Yuan, Shuo and Liu, Siqi and Zhang, Tong-Jie and Sun, Yan-Chun},
title = {{Four new observational $H(z)$ data from luminous red galaxies in the Sloan Digital Sky Survey data release seven}},
journal = {Research in Astronomy and Astrophysics}
}

@article{Moresco_2016,
doi = {https://doi.org/10.1088/1475-7516/2016/05/014},
year = {2016},
publisher = {},
volume = {2016},
number = {05},
pages = {014},
author = {Moresco, Michele and Pozzetti, Lucia and Cimatti, Andrea and Jimenez, Raul and Maraston, Claudia and Verde, Licia and Thomas, Daniel and Citro, Annalisa and Tojeiro, Rita and Wilkinson, David},
title = {{A 6\% measurement of the Hubble parameter at z $\sim$ 0.45: direct evidence of the epoch of cosmic re-acceleration}},
journal = {Journal of Cosmology and Astroparticle Physics}
}

@article{Simon_2005,
  title = {{Constraints on the redshift dependence of the dark energy potential}},
  author = {Simon, Joan and Verde, Licia and Jimenez, Raul},
  journal = {Phys. Rev. D},
  volume = {71},
  issue = {12},
  pages = {123001},
  numpages = {18},
  year = {2005},
  publisher = {American Physical Society},
  doi = {https://doi.org/10.1103/PhysRevD.71.123001}
}

@article{Stern_2010,
doi = {https://doi.org/10.1088/1475-7516/2010/02/008},
year = {2010},
publisher = {},
volume = {2010},
number = {02},
pages = {008},
author = {Daniel Stern and Raul Jimenez and Licia Verde and Marc Kamionkowski and S. Adam Stanford},
title = {{Cosmic chronometers: constraining the equation of state of dark energy. I: $H(z)$ measurements}},
journal = {Journal of Cosmology and Astroparticle Physics}
}

@article{Moresco_2012,
doi = {https://doi.org/10.1088/1475-7516/2012/08/006},
year = {2012},
publisher = {},
volume = {2012},
number = {08},
pages = {006},
author = {M. Moresco and A. Cimatti and R. Jimenez and L. Pozzetti and G. Zamorani and M. Bolzonella and J. Dunlop and F. Lamareille and M. Mignoli and H. Pearce and others},
title = {{Improved constraints on the expansion rate of the Universe
 up to z $\sim$ 1.1 from the spectroscopic evolution of cosmic chronometers}},
journal = {Journal of Cosmology and Astroparticle Physics}
}

@article{Moresco_2015,
    author = {Moresco, Michele},
    title = {{Raising the bar: new constraints on the Hubble parameter with cosmic chronometers at z $\sim$ 2}},
    journal = {Monthly Notices of the Royal Astronomical Society: Letters},
    volume = {450},
    number = {1},
    pages = {L16-L20},
    year = {2015},
    doi = {https://doi.org/10.1093/mnrasl/slv037}
}

@article{Ratsimbazafy_2017,
    author = {Ratsimbazafy, A. L. and Loubser, S. I. and Crawford, S. M. and Cress, C. M. and Bassett, B. A. and Nichol, R. C. and Väisänen, P.},
    title = {{Age-dating luminous red galaxies observed with the Southern African Large Telescope}},
    journal = {Monthly Notices of the Royal Astronomical Society},
    volume = {467},
    number = {3},
    pages = {3239-3254},
    year = {2017},
    doi = {https://doi.org/10.1093/mnras/stx301}
}

@article{Aghanim_2020,
	author = {Aghanim, N. and Akrami, Y. and Ashdown, M. and Aumont, J. and Baccigalupi, C. and Ballardini, M. and Banday, A. J. and Barreiro, R. B. and Bartolo, N. and Basak, S. and others},
	title = {{Planck 2018 results - VI. Cosmological parameters}},
	doi={https://doi.org/10.1051/0004-6361/201833910},
	journal = {A \& A},
	year = {2020},
	volume = {641},
	pages = {A6},
    publisher={EDP sciences}
}

@ARTICLE{Akaike_1974,
  author={Akaike, H.},
  journal={IEEE Transactions on Automatic Control}, 
  title={{A new look at the statistical model identification}}, 
  year={1974},
  volume={19},
  number={6},
  pages={716-723},
  doi={https://doi.org/10.1109/TAC.1974.1100705}
  }

@article{Schwarz_1978,
 author = {Gideon Schwarz},
 journal = {The Annals of Statistics},
 number = {2},
 pages = {461--464},
 publisher = {Institute of Mathematical Statistics},
 title = {{Estimating the Dimension of a Model}},
 urldate = {2026-06-20},
 volume = {6},
 year = {1978}
}

@article{Rezaei_2021,
  title={{Comparison between different methods of model selection in cosmology}},
  author={Rezaei, Mehdi and Malekjani, Mohammad},
  journal={The European Physical Journal Plus},
  volume={136},
  number={2},
  pages={219},
  year={2021},
doi={https://doi.org/10.1140/epjp/s13360-021-01200-w},
  publisher={Springer}
}

@article{Burnham_2004,
author = {Kenneth P. Burnham and David R. Anderson},
title ={{Multimodel Inference: Understanding AIC and BIC in Model Selection}},
journal = {Sociological Methods \& Research},
volume = {33},
number = {2},
pages = {261-304},
year = {2004},
doi = {https://doi.org/10.1177/0049124104268644}
}

@article{Kass_1995,
author = {Robert E. Kass and Adrian E. Raftery},
title = {{Bayes Factors}},
journal = {Journal of the American Statistical Association},
volume = {90},
number = {430},
pages = {773--795},
year = {1995},
publisher = {Taylor \& Francis},
doi = {https://doi.org/10.1080/01621459.1995.10476572}
}

@article{Visser_2004,
doi = {https://doi.org/10.1088/0264-9381/21/11/006},
year = {2004},
publisher = {},
volume = {21},
number = {11},
pages = {2603},
author = {Matt Visser},
title = {{Jerk, snap and the cosmological equation of state}},
journal = {Classical and Quantum Gravity},
}

@article{Bhagat_2025,
title = {{Exploring the viability of $f(Q,T)$ gravity: Constraining parameters with cosmological observations}},
journal = {Physics of the Dark Universe},
volume = {49},
pages = {102048},
year = {2025},
doi = {https://doi.org/10.1016/j.dark.2025.102048},
author = {Rahul Bhagat and Santosh V. Lohakare and B. Mishra}
}

@article{Mishra_2025,
    author = {Mishra, Sai Swagat and Kavya, N S and Sahoo, P K and Harko, Tiberiu},
    title = {{Padé cosmography and its insights into teleparallel gravity}},
    journal = {Monthly Notices of the Royal Astronomical Society},
    volume = {543},
    number = {3},
    pages = {2816-2835},
    year = {2025},
    doi = {https://doi.org/10.1093/mnras/staf1492}
}

@article{Pourojaghi_2025,
    author = {Pourojaghi, Saeed and Malekjani, Mohammad and Davari, Zahra},
    title = {{$\Lambda$CDM model against cosmography: a possible deviation after DESI 2024}},
    journal = {Monthly Notices of the Royal Astronomical Society},
    volume = {537},
    number = {1},
    pages = {436-447},
    year = {2025},
    doi = {https://doi.org/10.1093/mnras/staf037}
}

@article{Singh_2027,
author = {Singh, Kshetrimayum Govind and Singh, Kangujam Priyokumar},
title = {{Exploring late-time cosmic acceleration through parametrized deceleration parameter in $f(Q,L_m)$ gravity}},
journal = {International Journal of Geometric Methods in Modern Physics},
volume = {24},
number = {06},
pages = {2650203},
year = {2027},
doi = {https://doi.org/10.1142/S0219887826502038}
}

@article{Sahni_2003,
  title={{Statefinder—a new geometrical diagnostic of dark energy}},
  author={Sahni, Varun and Saini, Tarun Deep and Starobinsky, Alexei A and Alam, Ujjaini},
  journal={Journal of Experimental and Theoretical Physics Letters},
  volume={77},
  number={5},
  pages={201--206},
  year={2003},
  doi ={https://doi.org/10.1134/1.1574831},
  publisher={Springer}
}

@article{Alam_2003,
    author = {Alam, Ujjaini and Sahni, Varun and Deep Saini, Tarun and Starobinsky, A. A.},
    title = {{Exploring the expanding Universe and dark energy using the statefinder diagnostic}},
    journal = {Monthly Notices of the Royal Astronomical Society},
    volume = {344},
    number = {4},
    pages = {1057-1074},
    year = {2003},
    doi = {https://doi.org/10.1046/j.1365-8711.2003.06871.x}
}

@article{Sahni_2008,
  title = {{Two new diagnostics of dark energy}},
  author = {Sahni, Varun and Shafieloo, Arman and Starobinsky, Alexei A.},
  journal = {Phys. Rev. D},
  volume = {78},
  issue = {10},
  pages = {103502},
  numpages = {11},
  year = {2008},
  publisher = {American Physical Society},
  doi = {https://doi.org/10.1103/PhysRevD.78.103502}
}

\end{document}